\documentclass[12pt,a4paper,dvips]{article}
\usepackage{a4p}
\usepackage{cite,mcite}
\usepackage{graphicx}
\usepackage{physics }
\usepackage{l3_title,ifthen}
\usepackage{rotating}
\usepackage{epsfig}

\journalname{Phys. Lett. B}
\date{December 9, 2003}

\usepackage{Lep}
\preprint{2003-086}
\Lep{2}

\newlength{\capindent}
\newlength{\capwidth}
\newlength{\figwidth}
\newcommand{\icaption}[2][!*!,!]{\hspace*{\cpaindent}%
  \begin{minipage}{\capwidth}
    \ifthenelse{\equal{#1}{!*!,!}}%
      {\caption{#2}}%
      {\caption[#1]{#2}}
  \end{minipage}}
\input{l3at183.def}

\begin{document}
\begin{titlepage}
\title{Measurement of Triple-Gauge-Boson Couplings \\ 
       of the W Boson at LEP \\}
\vspace*{-20mm}
\author{The L3 Collaboration}
\vspace*{-10mm}

%
% The abstract
%
\begin{abstract}
The  CP-conserving triple-gauge-boson couplings, 
$\giZ$, $\kg$, $\Lg$, $\gvZ$, $\kZ$ and $\LZ$
are measured using hadronic and semi-leptonic W-pair events 
selected in  629~pb$^{-1}$ of data collected at LEP with the L3 detector
at centre-of-mass energies between $189$ and $209~\GeV$.
The results are combined with 
previous L3 measurements based on data collected at lower 
centre-of-mass energies and with the results from single-W 
production and from events with a
single-photon and missing energy. 
Imposing the constraints $\kZ = \giZ - \tan^2\theta_{W}
(\kg-1)$  
and $\LZ = \Lg$, we obtain for the C and P conserving couplings the
results:     
\begin{center}
\parbox{9cm}
{
\begin{eqnarray*}
\giZ & = &  \phantom{+} 0.966\pm0.033  \mathrm{(stat.)}  \pm0.015 \mathrm{(syst.)} \\
\kg  & = &  \phantom{+} 1.013\pm0.066  \mathrm{(stat.)}  \pm0.026 \mathrm{(syst.)} \\
\Lg  & = &  -0.021\pm0.035  \mathrm{(stat.)}  \pm0.017 \mathrm{(syst.)}.
\end{eqnarray*}
}
\end{center}
Results from the analysis of fully leptonic W-pair decays are also given. 
All results are in agreement with the Standard Model expectations and
confirm the existence of self-couplings among electroweak gauge
bosons. 
\end{abstract}
\submitted

\end{titlepage}
%
%%%%%%%%%%%%%%%%%%%%%%%%%%%%%%%%%%%%%%%%%%%%%%%%%%%%%%%%%%%%%%%%%%%%%%%%%%%%%%%
% Introduction
%%%%%%%%%%%%%%%%%%%%%%%%%%%%%%%%%%%%%%%%%%%%%%%%%%%%%%%%%%%%%%%%%%%%%%%%%%%%%%%
%
\section{Introduction}
\label{sec:intro}

The non-Abelian structure of the electroweak
theory~\cite{standard_model} implies the existence of trilinear self
couplings among gauge bosons.  
The vertices $\gamma$WW and ZWW are accessible at LEP   
through W-pair, single-W and single-photon production~\cite{LEP2YRSM}.

To lowest order, 
three Feynman diagrams contribute to W-pair production: the
$s$-channel $\gamma$ and Z exchange and the $t$-channel
$\nu_{\e}$ exchange.
The $s$-channel diagrams contain the $\gamma$WW and ZWW vertices.  
The $\gamma$WW vertex
appears in one of the $t$-channel Feynman diagrams contributing to
single-W production, $\EE\rightarrow\PW\EN$; at LEP
centre-of-mass energies, $\sqrt{s}$, the contribution from the similar 
diagram containing the ZWW vertex is negligible.  
The $\gamma$WW vertex also contributes to the $\rm e^+e^- \rightarrow
\nu_{\mathrm{e}}\bar\nu_{\mathrm{e}}\gamma$ process through photon
production in W-boson fusion.

Assuming only Lorentz invariance, the most general form of the
$\gamma$WW and ZWW vertices is parametrised in terms 
of seven complex triple-gauge-boson couplings (TGC's) each~\cite{Hagiwara87}.  
Retaining only CP-conserving
couplings and assuming electromagnetic gauge invariance, six real TGC's
remain, namely $\giZ$, $\kg$, $\Lg$, $\gvZ$, $\kZ$ and $\LZ$.
At tree level within the Standard Model, $\giZ=\kg=\kZ=1$ and $\gvZ=\Lg=\LZ=0$.
Except $\gvZ$, these TGC's also conserve C and P separately.
The requirement of custodial $\mathrm{SU(2)}$ symmetry leads to the 
relations $\kZ = \giZ - \tan^2\theta_{W} (\kg-1)$ and $\LZ =
\Lg$~\cite{DKLDXYSchild,LEP2YRAC}, where $\theta_{W}$
is the weak mixing angle.  
When these constraints are applied,  
$\giZ$, $\kg$ and $\Lg$ correspond to the operators in a linear
realisation of a gauge-invariant effective Lagrangian that do not
affect the gauge-boson propagators at tree level~\cite{LEP2YRAC}.  
The $\giZ$, $\kg$ and $\Lg$ couplings are studied assuming these constraints.
The analysis is based on the study of multi-differential cross
sections measured in hadronic and semi-leptonic W-pair events. 
Measurements at lower $\sqrt{s}$~\cite{l3-189} are included, as 
well as events selected by the single-W analysis~\cite{l3-252} and
events with a single photon and missing energy~\cite{singlePhoton}.
Results from the analyses of fully leptonic W-pair decays are also given. 
Results on TGC's were also published by experiments at hadron
colliders~\cite{collidertgc} and at LEP~\cite{LEPtgc}.

%%%%%%%%%%%%%%%%%%%%%%%%%%%%%%%%%%%%%%%%%%%%%%%%%%%%%%%%%%%%%%%%%%%%%%%%%%%%%%%
% Data and MC samples
%%%%%%%%%%%%%%%%%%%%%%%%%%%%%%%%%%%%%%%%%%%%%%%%%%%%%%%%%%%%%%%%%%%%%%%%%%%%%%%
\section{Data and Monte Carlo Samples}
\label{sec:datamc}
 
The data sample collected by the L3 detector~\cite{l3_det} in the
years from 1998 through 2000 is used in the W-pair analysis. 
It corresponds to an integrated luminosity of
629.2~pb$^{-1}$ at $\sqrt{s}=189 -209~\GeV$, 
detailed in Table~\ref{tab:luminosities}.
An additional 76.4~pb$^{-1}$ of data at $\sqrt{s}=161 - 183~\GeV$ is used for the 
single-W analysis.

The following Monte Carlo
event generators are used to simulate the signal and background
reactions:  
KandY~\cite{kandy} and EXCALIBUR~\cite{EXCALIBUR}
for $\EEFFFFG$; 
PYTHIA~\cite{JETSET73} for $\EEQQG, \EE \ra \mathrm{ZZ}(\gamma)$ and
$\EE \ra \mathrm{Z}\EE$; 
KK2f~\cite{KK2f} for $\EEQQG, \EE \ra \MMG$ and  $\EE \ra \TTG$;
BHAGENE3 \cite{BHAGENE}, BHWIDE~\cite{BHWIDE} and TEEGG~\cite{TEEGG}
for $\EEEEG$ and 
DIAG36~\cite{DIAG36} and PHOJET~\cite{PHOJET} for lepton and hadron
production in two-photon collisions, respectively.
The KandY program, used to generate
W-pair events, combines the four-fermion generator KORALW~\cite{KORALW} with
the $\cal{O}(\alpha)$ radiative corrections in the leading-pole
approximation~\cite{DPA} implemented in the YFSWW program~\cite{YFSWW3}. 

The response of the L3 detector is modelled with the
GEANT~\cite{geant} program which includes 
effects of energy loss, multiple scattering and showering in the
detector materials and in the beam pipe.
Time-dependent detector inefficiencies, as monitored during the data
taking period, are included in the simulations.

%%%%%%%%%%%%%%%%%%%%%%%%%%%%%%%%%%%%%%%%%%%%%%%%%%%%%%%%%%%%%%%%%%%%%%%%%%%%%%%
% Event Selection
%%%%%%%%%%%%%%%%%%%%%%%%%%%%%%%%%%%%%%%%%%%%%%%%%%%%%%%%%%%%%%%%%%%%%%%%%%%%%%%
\section{Event Selection}
\label{sec:evtsel}

\subsection{W-pair Events}

The event selection is based on that described in
Reference~\citen{189paper} and its results are detailed in
Reference~\citen{wwxsecpaper}. 
The visible fermions in the final state are reconstructed as
electrons, muons, jets corresponding to
decay products of $\tau$ leptons, and hadronic jets corresponding to
quarks.
Only events containing leptons with an unambiguous charge assignment are retained. 
The numbers of selected hadronic, semi-leptonic and fully leptonic
W-pair events and
the expected background are given in Table~\ref{tab:events-all}. 

Kinematic fits are performed
to improve the resolution of the measured fermion energies and angles 
and to determine neutrino momenta in semi-leptonic events.
Four-momentum conservation and equal mass of the two W bosons are
imposed as constraints. 
In $\QQTN$ events, the energies of the two hadronic jets
are rescaled by a common factor so that their sum equals $\sqrt{s}/2$. 
The four jets in hadronic events are paired to form W bosons 
by a neural network based on
the difference and sum of the masses of the jet pairs, the sum and the
minimum of the angles between paired jets, the energy
difference between the jet pairs and between the paired jets,
the value of the matrix element for the process $\EE \ra \PW^+\PW^- \ra ffff$
as calculated with EXCALIBUR from the jet four-momenta, and
the difference between the charges of the jet pairs as determined from the jet 
charges~\cite{l3-189}. The correct pairing is found 
for 77\% of the selected Monte Carlo events.

\subsection{Single-W Events}

The $\EE\rightarrow\PW\EN$ process typically has an electron scattered
at very low polar angle, so that only the decay products of the W
boson are observed as single-lepton events or acoplanar
jets. Single-lepton events are selected by exploiting their peculiar
signature in the detector, while a neural network is used to isolate
hadronic single-W events from the background~\cite{l3-252}.
The hadronic sample consists of 740 events out of which 156
are also accepted by the semi-leptonic W-pair selections.
From Monte Carlo studies, about $75\%$ of this overlap consists of 
W-pair events, mostly $\QQTN$ events, while only $7\%$ consists of 
single-W events, the remainder being $\EE \ra \QQG$ events.
In order to avoid double counting, these events are
considered in the W-pair sample only.  

The numbers of selected single-W events 
and the expected background, after the
removal of the overlapping events, are reported in
Table~\ref{tab:events-all}. 

%%%%%%%%%%%%%%%%%%%%%%%%%%%%%%%%%%%%%%%%%%%%%%%%%%%%%%%%%%%%%%%%%%%%%%%%%%%%%%%
%Event Reconstruction
%%%%%%%%%%%%%%%%%%%%%%%%%%%%%%%%%%%%%%%%%%%%%%%%%%%%%%%%%%%%%%%%%%%%%%%%%%%%%%%
\section{Event Reconstruction}
\label{sec:phasespace}

% \subsection{W-pair Analysis}

For unpolarised initial states, summing over final-state fermion
helicities, fixing the mass of the W 
boson and neglecting photon radiation, five angles
completely describe the four-fermion final state originating 
from W-pair decay.  
These angles are the production angle of
the W$^-$ boson, $\Theta_{\PW^-}$, and the polar and the azimuthal decay
angles of the fermion in W$^-$ decays and the anti-fermion in W$^+$ decays,
calculated in the rest frame of the W boson.
TGC's affect the total production cross section, the
W production angle, and the
polarisations of the two W bosons, which in turn determine the 
W decay angles.

\begin{figure}[p]
\begin{center}
\epsfig{file=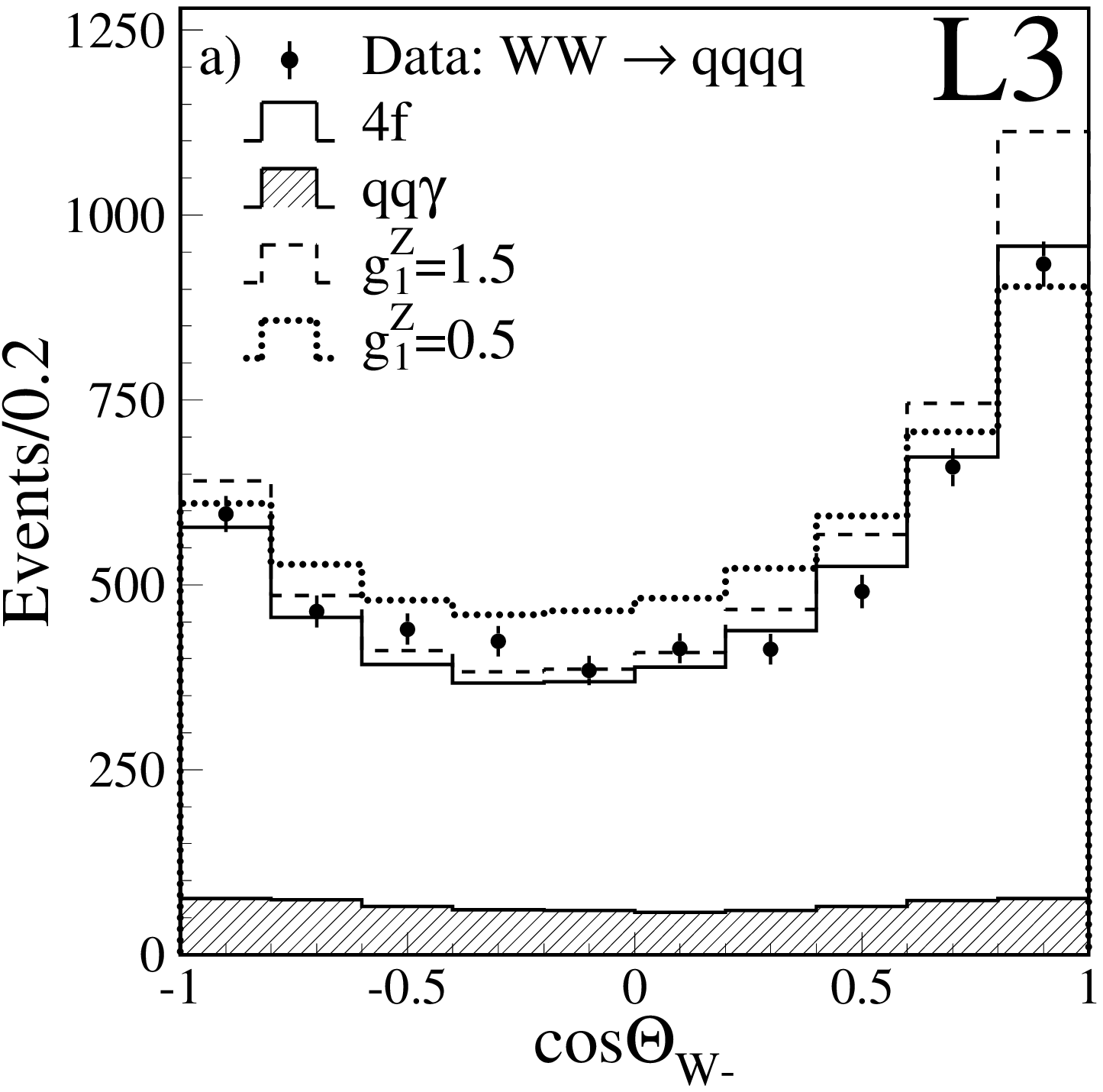,width=0.49\linewidth}\hfill %\hspace*{1.5cm}
\epsfig{file=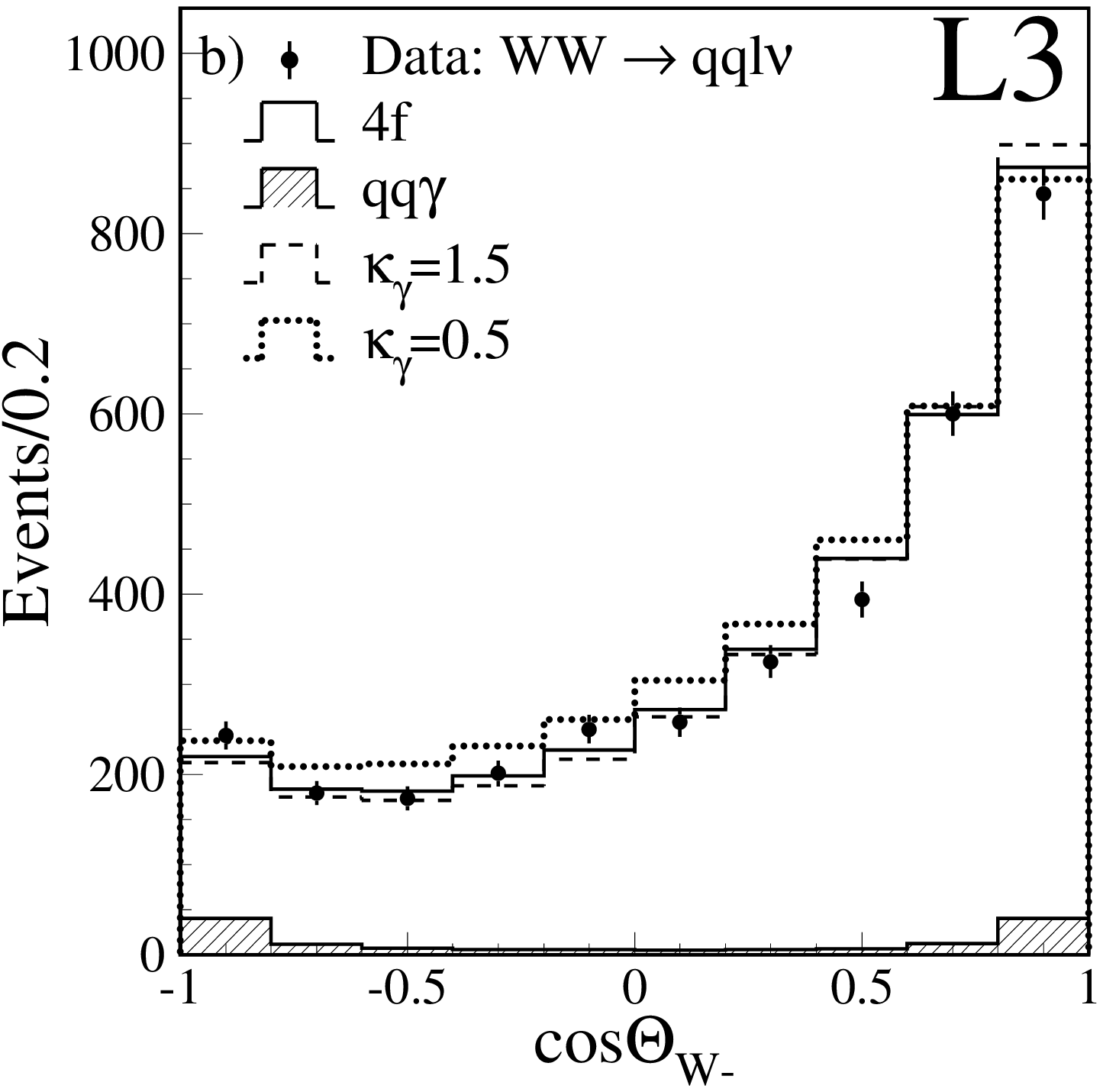,width=0.49\linewidth}\\
\caption[]{Distributions of the reconstructed W$^-$ production angle,
  $\cos\Theta_{\PW^-}$, in 
  a) hadronic and b) semi-leptonic W-pair events. 
  Data are shown, together
  with the expectations for the Standard Model and for anomalous 
  values of TGC's.}  
\label{fig:wwctw}
\end{center}
\end{figure}

For semi-leptonic W-pair events, the W$^-$ production angle is
reconstructed from the hadronic part of the event, and the sign
of $\cos\Theta_{\PW^-}$ is
determined from the lepton charge.
If both W bosons decay into hadrons, 
the W charge assignment follows from jet-charge technique~\cite{l3-189}.
This charge assignment is found to be correct for
69\% of Monte Carlo events with correctly paired jets.  
The distributions of  $\cos \Theta_{\PW^-}$
for hadronic and semi-leptonic events are shown in
Figure~\ref{fig:wwctw} where, 
for illustrative purposes, all data are combined.

\begin{figure}[p]
\begin{center}
\epsfig{file=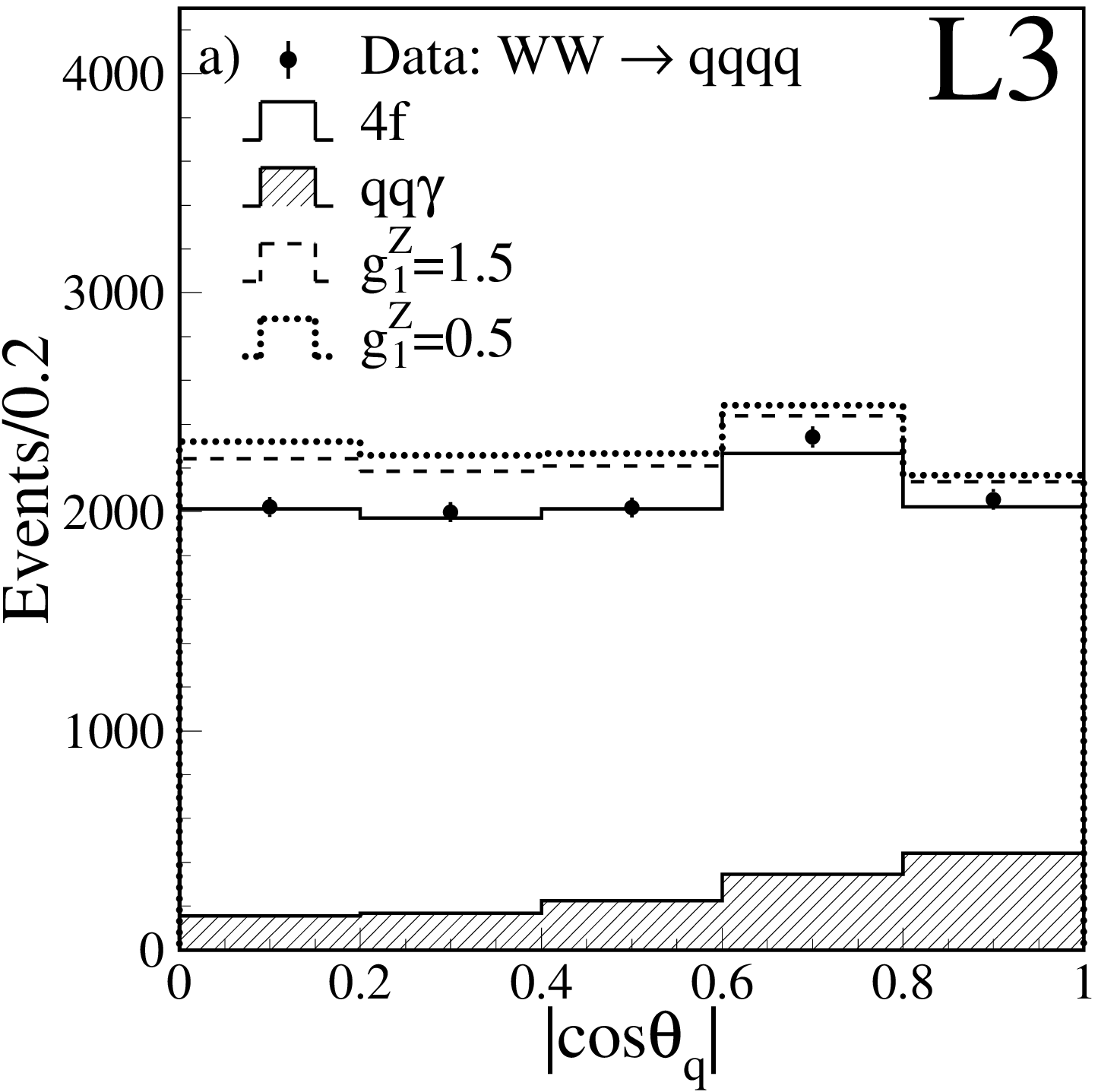,width=0.49\linewidth}\hfill %\hspace*{1.5cm}
\epsfig{file=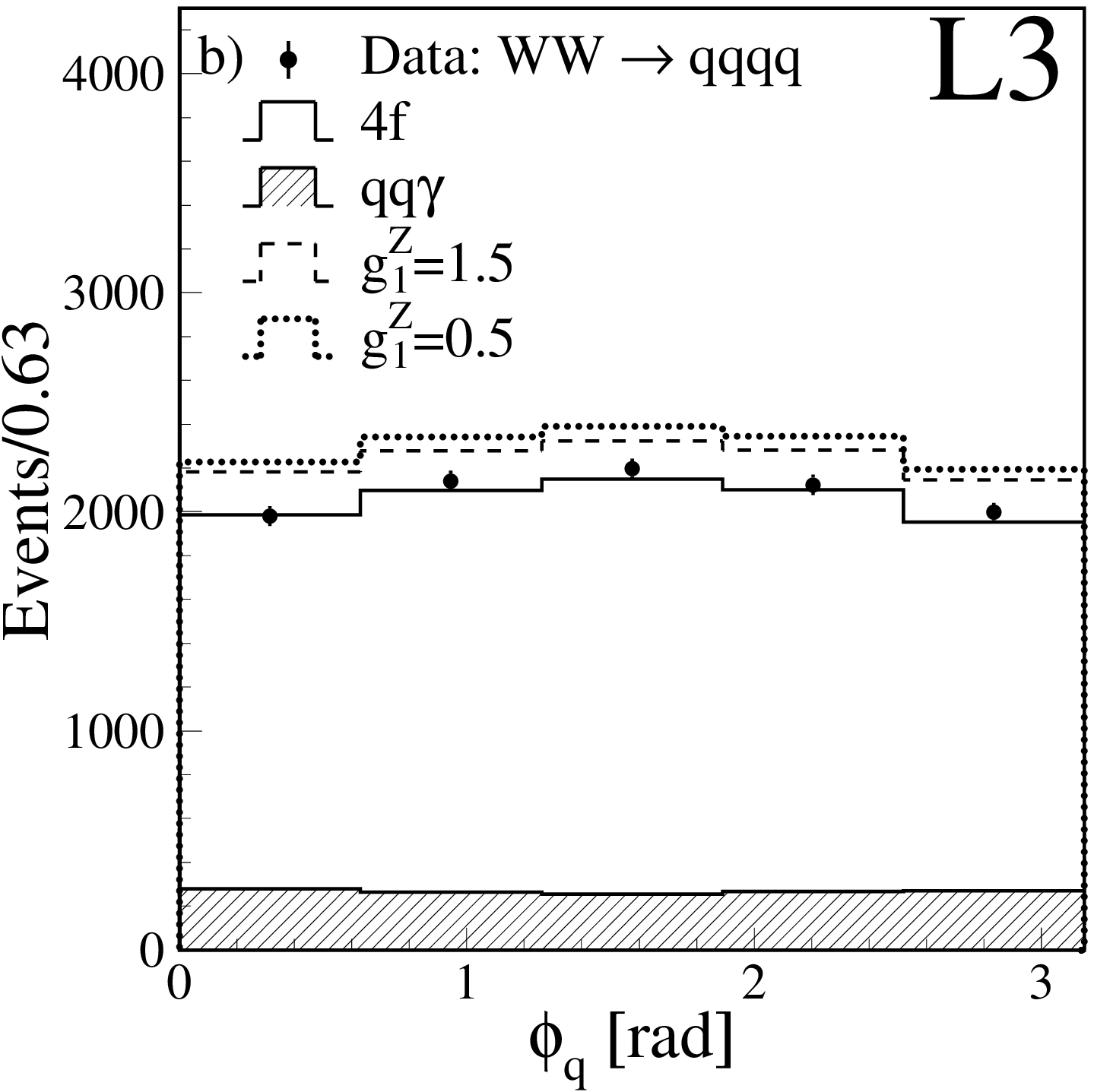,width=0.49\linewidth}\\
\caption[]{Distributions of the reconstructed W decay angles in
  hadronic W-pair events, a) $| \cos\theta_q|$  and b) $\phi_q$.  
  Distributions for  W$^+$ and W$^-$ bosons are 
  combined.
  Data are shown, together with the
  expectations for the Standard Model and for anomalous values
  of the TGC's.}
\label{fig:wwdecayqqqq}
\end{center}
\end{figure}

\begin{figure}[p]
\begin{center}
\epsfig{file=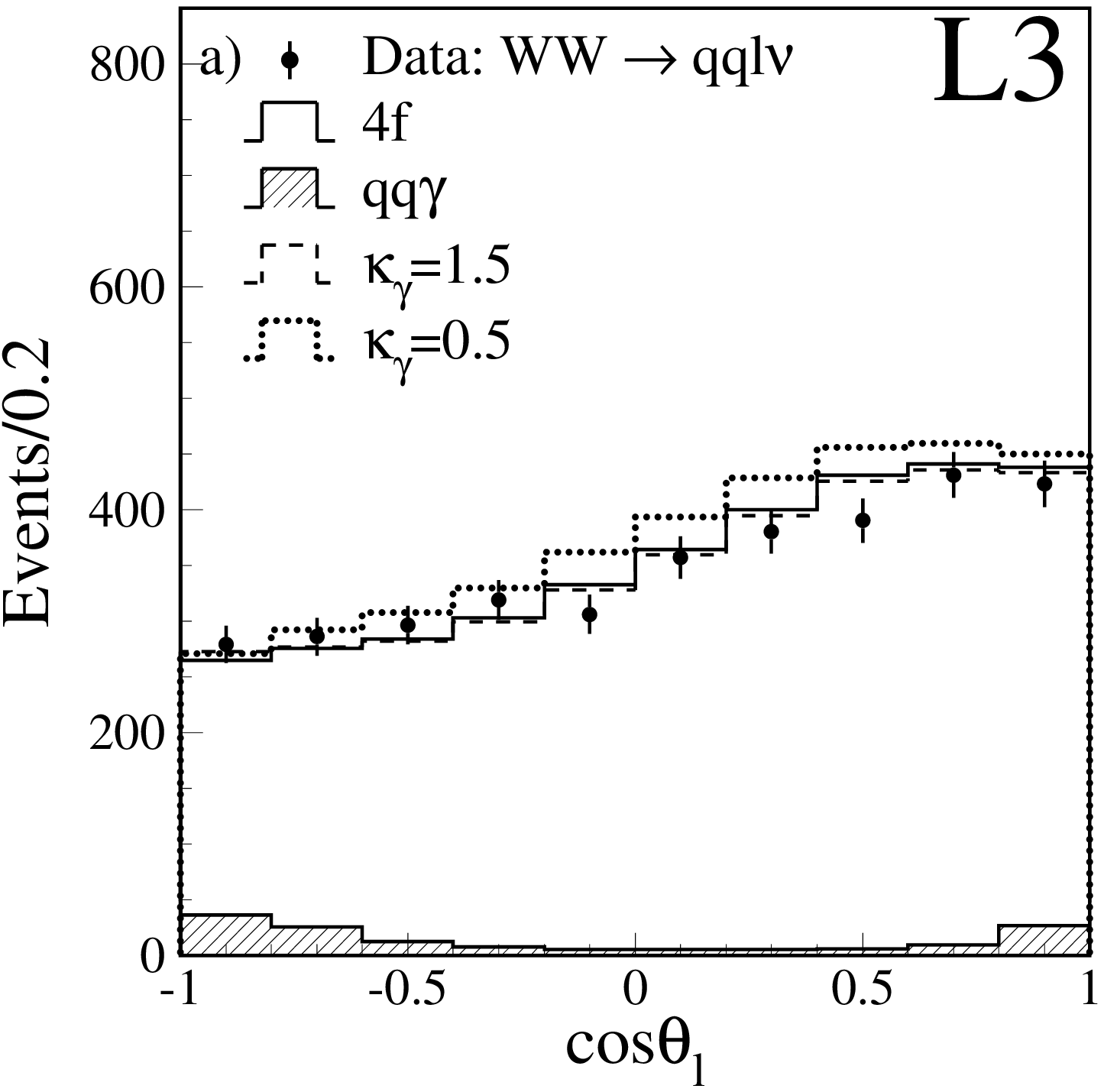,width=0.49\linewidth}\hfill %\hspace*{1.5cm}
\epsfig{file=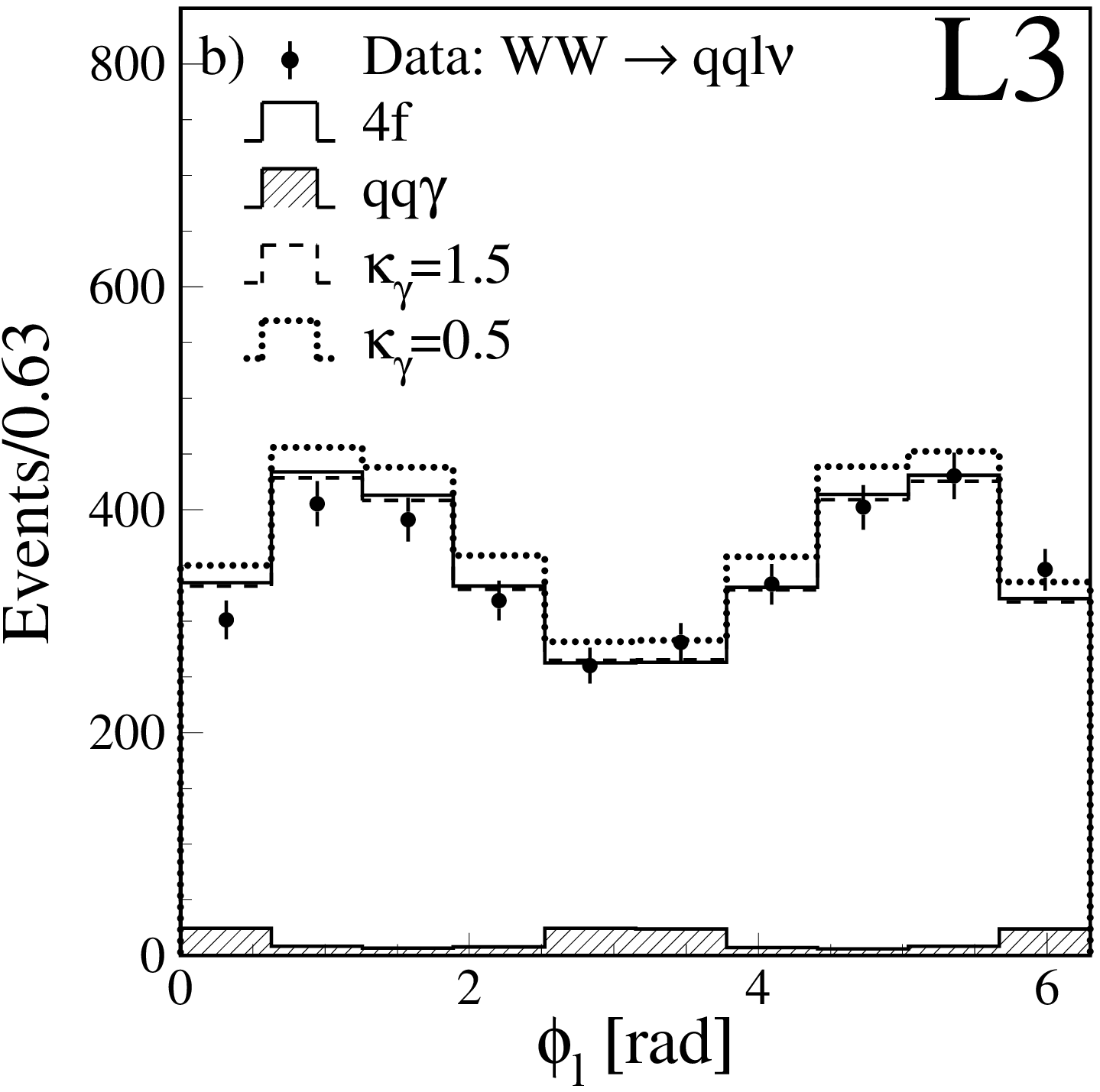,width=0.49\linewidth}\\
\vskip 0.5cm
\epsfig{file=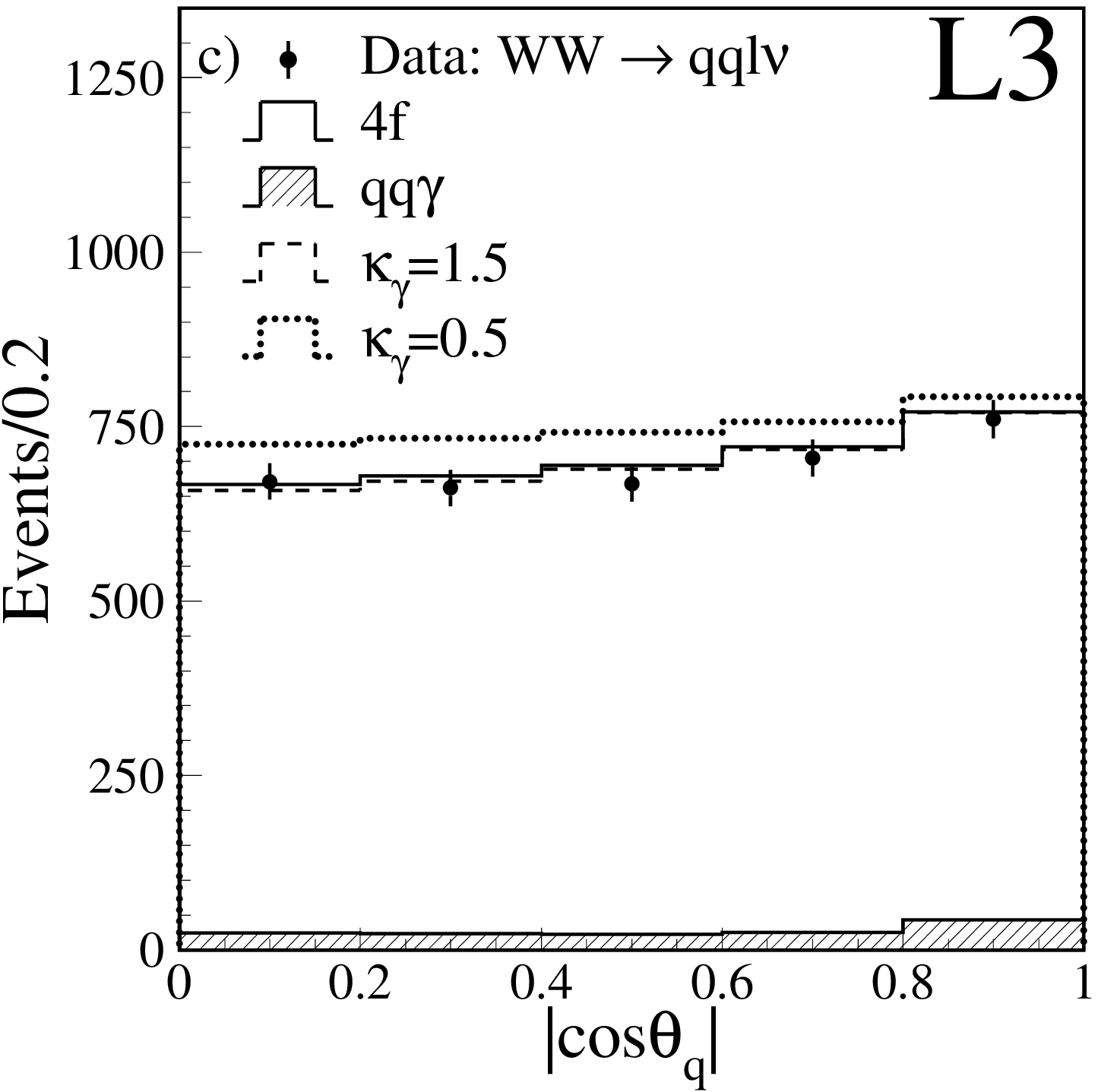,width=0.49\linewidth}\hfill %\hspace*{1.5cm}
\epsfig{file=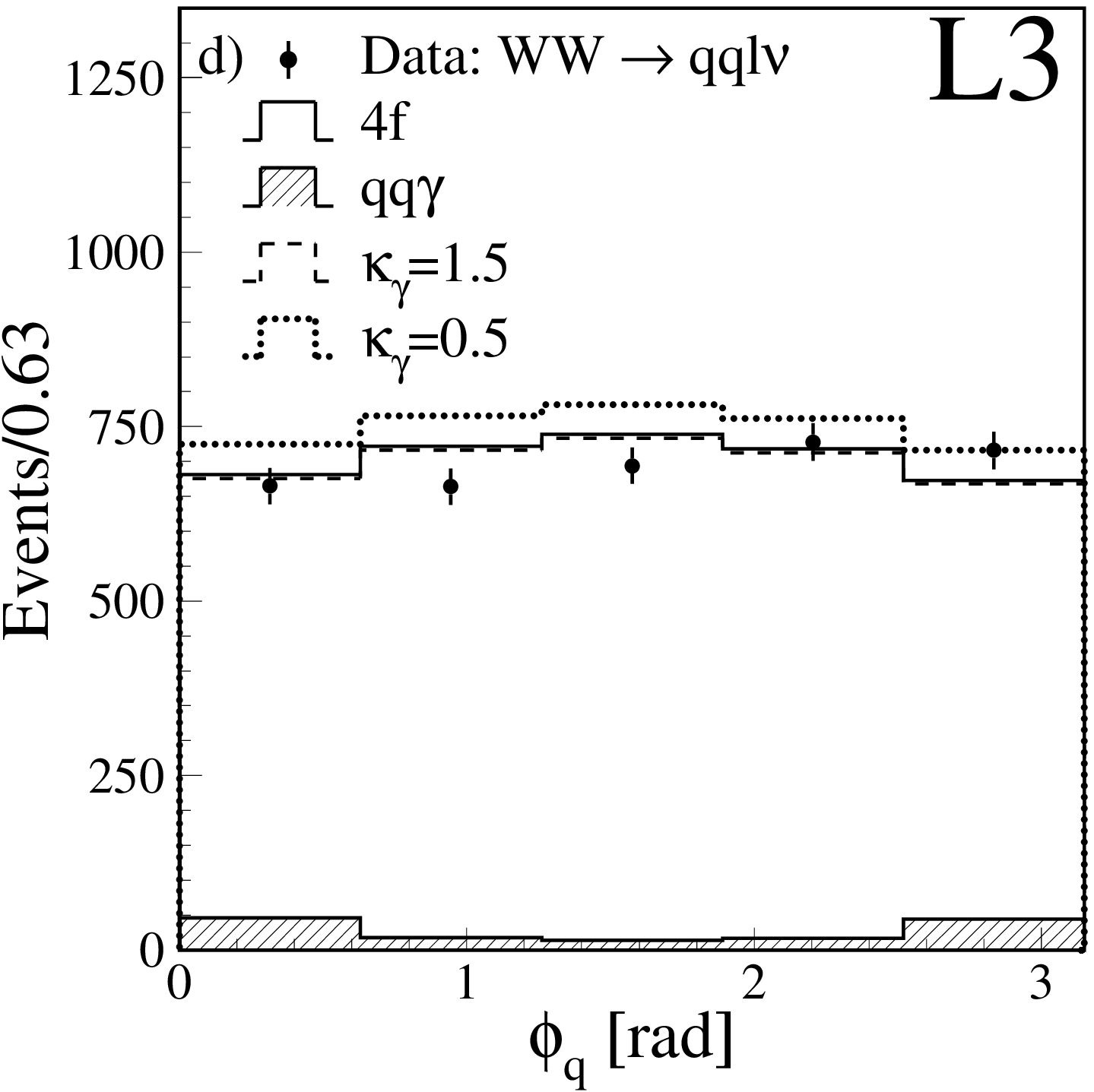,width=0.49\linewidth}\\
\caption[]{Distributions of the reconstructed W decay angles in
  semi-leptonic events: the production angles of the lepton, 
  a) $\cos\theta_{\ell}$ and b) $\phi_\ell$, and
  the decay angles of W bosons decaying into hadrons,  
  c) $|\cos\theta_q|$ and d) $\phi_q$. 
  Data are shown, together with the
  expectations for the Standard Model and for anomalous  values of the TGC's. }
\label{fig:wwdecayqqln}
\end{center}
\end{figure}

The charge of the lepton allows the 
reconstruction of the decay angles $\theta_\ell$ and $\phi_\ell$.
Jet-charge
determination is not adequate to determine the quark charge and
a two-fold ambiguity arises for the decay angles
of W bosons decaying into hadrons, 
$(\cos\theta_q,\phi_q)\leftrightarrow(-\cos\theta_q,\pi+\phi_q)$.  
The $\phi_q$ distribution is restricted to the interval
$(0,\pi]$ and the jet with $\phi_q \in (0,\pi]$ is assigned to the quark
or the anti-quark originating from the decay of W$^-$ or W$^+$
respectively. 
The absolute value of the cosine of the polar decay angle is considered.
The distributions of the hadronic decay angles for the hadronic
channel and the leptonic and hadronic decay
angles for the semi-leptonic channels are shown in
Figures~\ref{fig:wwdecayqqqq} and \ref{fig:wwdecayqqln},
respectively.  

% lnln
Fully leptonic W-pair decay channels with final state
muons and electrons are also analysed. The presence of two neutrinos
prevents an unambiguous reconstruction of the event. Assuming no 
initial-state radiation, and fixing the mass of the W boson, the production angle
of the latter is kinematically derived with a two-fold
ambiguity~\cite{LEP2YRAC}. Due to resolution effects, about 40\% of the
events yield complex solutions and are not considered. A weight of one
half is given to each solution of the retained events.

%%%%%%%%%%%%%%%%%%%%%%%%%%%%%%%%%%%%%%%%%%%%%%%%%%%%%%%%%%%%%%%%%%%%%%%%%%%%
%Data Analysis
%%%%%%%%%%%%%%%%%%%%%%%%%%%%%%%%%%%%%%%%%%%%%%%%%%%%%%%%%%%%%%%%%%%%%%%%%%%%
\section{Data Analysis}
\label{sec:fit}

\subsection{Fit Method}

\begin{figure}[p]
\begin{center}
\epsfig{file=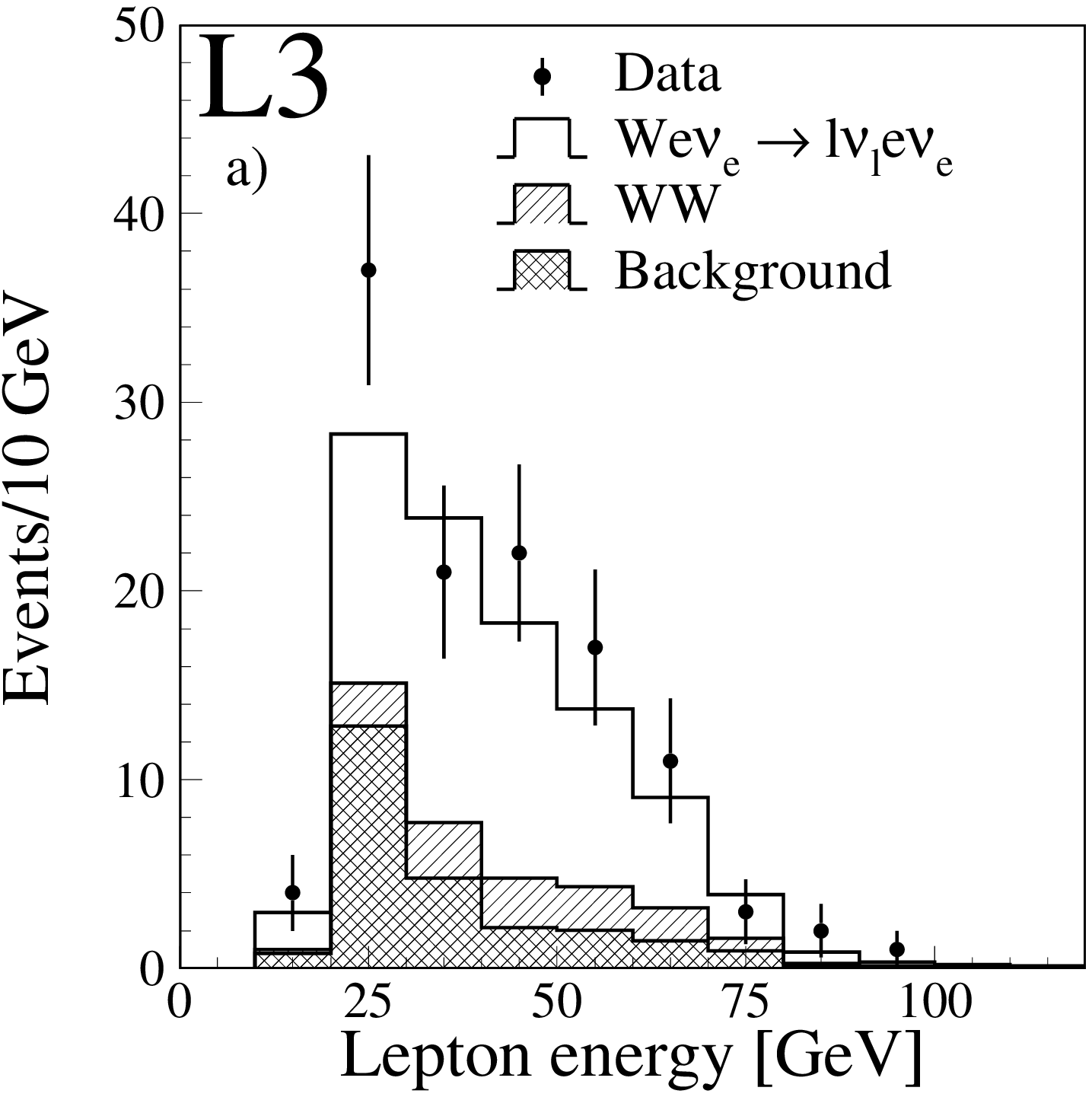,width=0.49\linewidth}\hfill
\epsfig{file=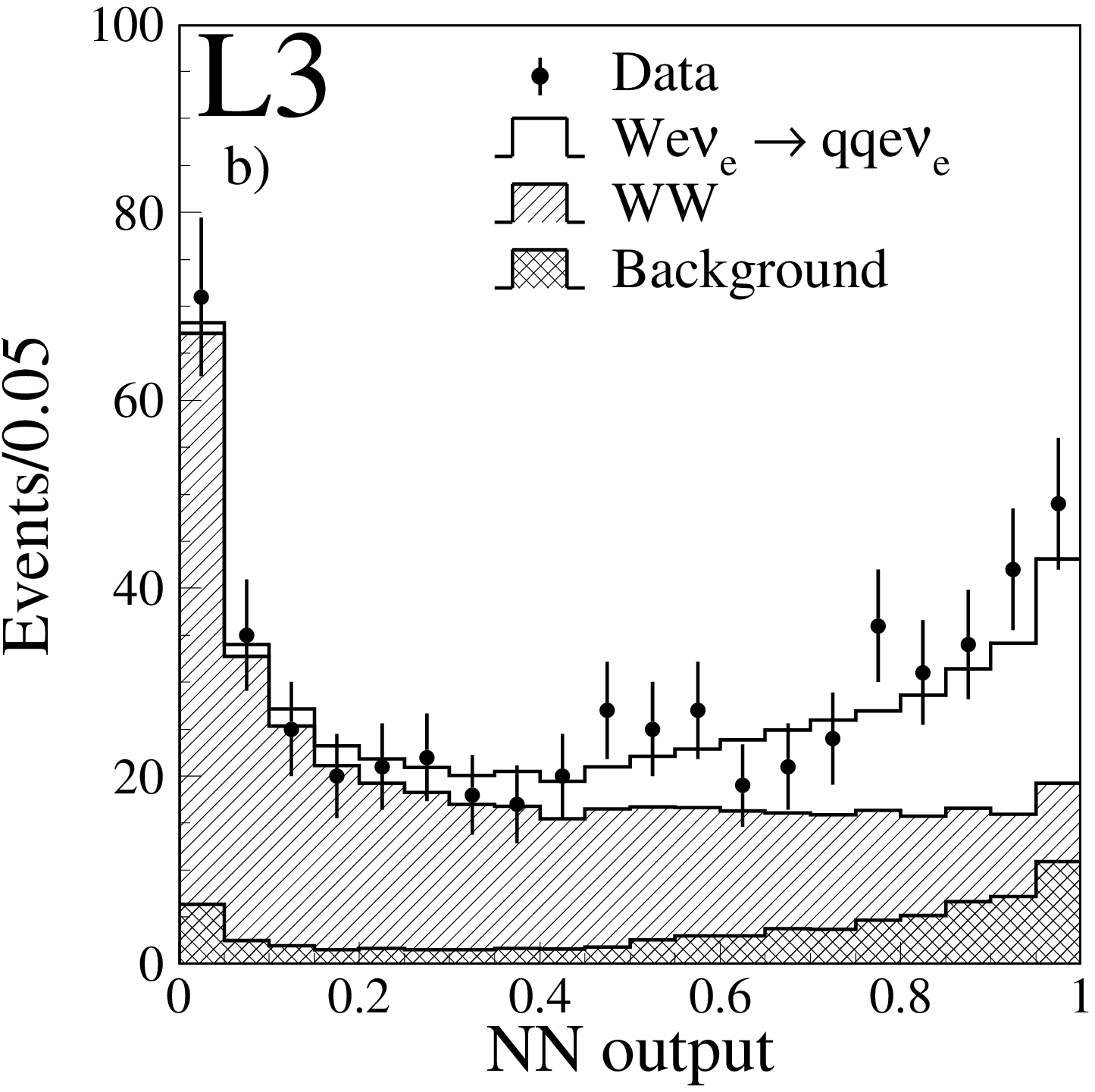,width=0.49\linewidth}\\
\caption[]{Distribution of a) the energy spectrum of the lepton in leptonic single-W 
  events and b) the output
  of the neural network used in
  the selection of hadronic single-W events.
}
\label{fig:swdist}
\end{center}
\end{figure}

Binned maximum likelihood fits are used to perform the TGC
measurement. Bin sizes are chosen so as to optimise sensitivity for the
given Monte Carlo statistics.
For hadronic and semi-leptonic W-pairs, the likelihoods depend on
the W production and decay angles. 
For $\cos \Theta_{\PW^-}$, 12 bins are considered in the
hadronic channel, 10 bins for $\QQEN$ and $\QQMN$ events and 8 bins 
in the $\QQTN$ channel.
For the leptonic decay angles  $\cos \theta_{\ell}$ and $\phi_{\ell}$,
4 bins are used, while 3 bins are considered for the hadronic
decay angles $|\cos \theta_{q}|$ and $\phi_{q}$. 
For leptonic single-W events, the lepton energy is used in the fit, 
with bins of $5 \gev$. Its distribution is  
shown in Figure~\ref{fig:swdist}a.
For hadronic single-W events
the neural network output, whose distribution is shown in
Figure~\ref{fig:swdist}b, is used in the fit. 
It is divided in bins of 0.01.

For each decay channel and value of $\sqrt{s}$, the likelihood
is defined as the product of the Poisson probabilities of occupation
in each bin of the phase space as a function of a given set of
couplings $\Psi$:  
\begin{eqnarray}
  L(\Psi) & = &  \prod_i^{\mathrm{bins}}  
  \frac{e^{-\mu_i(\Psi)} \mu_i(\Psi)^{N_i}}
       {N_i !}\; , 
\end{eqnarray}
where $\mu_i$ is the expected number of signal and background 
events in the $i$-th bin
and $N_i$ is the corresponding observed number of events. 
The dependence of $\mu_i$ on $\Psi$ is determined 
by a generator level reweighting procedure 
applied to fully simulated Monte Carlo events.
For any value of $\Psi$, 
the weight $R$ of the $n$-th event 
generated with TGC value $\Psi_{\gen}$ is:
\begin{eqnarray}
R(\Omega_n,\Psi,\Psi_{\gen}) & = & 
\frac
{\left|{\cal M}(\Omega_n,\Psi)\right|^2}
{\left|{\cal M}(\Omega_n,\Psi_{\gen})\right|^2}\,,
\label{eqn:weight}
\end{eqnarray}
where ${\cal M}$ is the matrix element of the final
state considered, evaluated~\cite{EXCALIBUR} for the generated phase space $\Omega_n$,
which includes radiated photons.  

The expected number of events in the $i$-th bin is:
\begin{eqnarray}
  \mu_{i}(\Psi) & = & 
  \sum_l^{\mathrm{sig+bg}} \left(
  \frac{\sigma_l^{\gen} \; \cal{L}}{N_l^{\gen}}
  \sum_{j}^{n_i} R_l(\Omega_j,\Psi,\Psi_{\gen})
  \right)  \,,
\end{eqnarray}
where the first sum runs over all signal and background samples, and
$\sigma_l^{\gen}$ 
denotes the cross section corresponding to
the total Monte Carlo sample containing $N_l^{\gen}$  
events and $\cal{L}$ is the integrated luminosity.
The second sum extends over the number $n_i$ of accepted Monte Carlo
events in the $i$-th bin.  
This definition takes properly into account detector effects 
and $\Psi$-dependent efficiencies and purities.  
For background sources which are independent of TGC's, $R_l =1$.
The fitting method described above determines the TGC's without any
bias as long as the Monte Carlo correctly describes photon radiation and
detector effects such as resolution and acceptance functions.  
Different channels and centre-of-mass
energies are combined by multiplying together the corresponding likelihoods.

The following results are obtained for hadronic and semi-leptonic W-pairs 
and for their combination, allowing one coupling to vary
while fixing the others to their Standard Model
values: 
\begin{eqnarray}
\!\!\!\!\!  \giZ =  \phantom{+}0.914\apm{0.065}{0.056} \; (\QQQQ) & \; \; \;
  \giZ =  \phantom{+}0.974\apm{0.039}{0.038} \; (\QQLN) & \; \; \;
  \giZ =  \phantom{+}0.959\apm{0.034}{0.033} \; (\mathrm{combined})
  \nonumber \\  
\!\!\!\!\!  \kg = \; \; \phantom{+}0.89\apm{0.12}{0.10} \; \; (\QQQQ) & \; \; \;   
  \kg =   \phantom{+}0.918\apm{0.097}{0.085} \; (\QQLN) & \; \; \;
  \kg =   \phantom{+}0.907\apm{0.074}{0.067} \; (\mathrm{combined})
  \nonumber \\  
\!\!\!\!\!  \Lg =    -0.102\apm{0.069}{0.058} \; (\QQQQ) & \; \; \;   
  \Lg =    -0.026\apm{0.040}{0.038} \; (\QQLN) & \; \; \;
  \Lg =    -0.044\apm{0.036}{0.033} \; (\mathrm{combined}) \, .
  \nonumber 
\end{eqnarray}
These couplings are determined under the constraints  $\kZ = \giZ -
\tan^2\theta_{W} (\kg-1)$ and $\LZ = \Lg$. Relaxing these
constraints, and fixing all other couplings to their Standard Model
values, yields: 
\begin{eqnarray}
\!\!\!\!\!  \gvZ =  \; \; \phantom{+}0.20\apm{0.21}{0.22} \; \; (\QQQQ) & \; \; \;
  \gvZ =  \; \; -0.10\apm{0.17}{0.17} \; \; (\QQLN) & \; \; \;
  \gvZ =  \; \; \phantom{+}0.00\apm{0.13}{0.13} \; \; (\mathrm{combined})
  \nonumber \\  
\!\!\!\!\!  \kZ =   \phantom{+}0.856\apm{0.108}{0.091} \; (\QQQQ) & \; \; \;   
  \kZ =   \phantom{+}0.957\apm{0.068}{0.066} \; (\QQLN) & \; \; \;
  \kZ =   \phantom{+}0.921\apm{0.059}{0.056} \; (\mathrm{combined})
  \nonumber \\  
\!\!\!\!\!  \LZ =   -0.179\apm{0.108}{0.085} \; (\QQQQ) & \; \; \;   
  \LZ =   -0.038\apm{0.066}{0.063} \; (\QQLN) & \; \; \;
  \LZ =   -0.070\apm{0.060}{0.057} \; (\mathrm{combined})\, .
  \nonumber 
\end{eqnarray}
The fit to fully leptonic W-pair events yields:
\begin{eqnarray}
  \; \; \; \; \; \; \; \; \; \; \; \; \; \; \; \; \; \; \; \; \; \; \; \;
  \giZ  =  0.91\apm{0.22}{0.16}  \; \; \;\; \; \; &
  \kg   =  1.07\apm{0.61}{0.38}  \; \; \;\; \; \;  &
  \Lg   =  -0.16\apm{0.15}{0.12}
  \nonumber
\end{eqnarray}
Due to the large statistical uncertainties of this channel,
compared to the other W-pair decay channels, these
results are not considered in the following combinations.

\subsection{Cross Checks}

The fitting procedure is tested to high
accuracy by fitting large Monte Carlo samples, typically a hundred
times the  size of the data.
TGC values are varied in a range corresponding to three
times the expected statistical uncertainty 
and are correctly reproduced by
the fit~\cite{villa,dierckxsens}.

The fit results are found to be independent of the value
$\Psi_{\gen}$ of the Monte Carlo sample subjected to the reweighting
procedure. 

The statistical uncertainties given by the fit are
tested by fitting, for each final state, several hundreds of small Monte
Carlo samples of the size of the data samples.  The width of the
distribution of the fitted central values agrees well with the mean of
the distribution of the uncertainties.

An independent analysis, based on
optimal observables technique~\cite{Fanourakisneutraltgc}, is
performed for the W-pair events and used as a cross check.
Both the central values and the uncertainties agree  
with those from the binned maximum likelihood
fit.

\subsection{Single-Photon Events}

Single-photon events
are mainly due to initial state radiation (ISR) 
in neutrino-pair production through $s$-channel
Z-boson exchange or $t$-channel W-boson exchange.  A small fraction of
events is due to W-boson fusion through the WW$\gamma$ vertex, which
gives access to $\kappa_{\gamma}$ and $\lambda_{\gamma}$.  Data at $\sqrt{s}=189-209~\GeV$ are
analysed~\cite{singlePhoton} and 1898 events are selected while 1905
are expected from the Standard Model. The KK2f Monte Carlo
program~\cite{KK2f} is used to simulate the $\rm e^+ e^-
\rightarrow \nu \bar{\nu}\gamma$ process and effects of TGC's are obtained by a
reweighting procedure~\cite{was_tgc}.

Binned maximum likelihood fits to the photon energy and polar angle
yield the results given in Table~\ref{tab:res1d1}.
The systematic uncertainties are dominated by uncertainties on the selection
efficiency~\cite{singlePhoton}, on the cross section~\cite{KK_theory}
and on the TGC modelling~\cite{was_highqed}.

%%%%%%%%%%%%%%%%%%%%%%%%%%%%%%%%%%%%%%%%%%%%%%%%%%%%%%%%%%%%%%%%%%%%%%%%%%%%
%Systematic Uncertainties
%%%%%%%%%%%%%%%%%%%%%%%%%%%%%%%%%%%%%%%%%%%%%%%%%%%%%%%%%%%%%%%%%%%%%%%%%%%%
\section{Systematic Uncertainties}
\label{sec:syst}

The systematic uncertainties for
W-pair events are summarised in Table~\ref{tab:sysww}.
The largest contributions are due to the limited Monte Carlo
statistics and to uncertainties on the background modelling, 
the W-pair cross section and the lepton charge reconstruction.

Systematic effects typically induce a shift in the position of the
maximum of the likelihood as well as a change of sensitivity.  For
sources of systematic uncertainties evaluated by varying a parameter
between two extremes of a range, if the sensitivity loss is larger
than the gain, the total uncertainty is evaluated as the sum in
quadrature of the difference between the loss and the gain and of the
shift in the maximum of the likelihood.  If the gain in sensitivity is
larger than the loss, only the shift in the maximum is quoted as
systematic uncertainty.

An uncertainty of 0.5\% on the $\EE\rightarrow \PW^+\PW^-$
cross section is assumed~\cite{MC-ffff}, based on the
predictions of KandY and RacoonWW~\cite{RACOONWW}. 
Both programs use either the leading-pole or 
the double-pole approximation. 
The $\cos\Theta_{\mathrm{W}^-}$ distribution expected for
these $\mathcal{O}(\alpha)$ calculations are compared and found to 
agree, in average slope, up to $0.4\%$. This value is assigned as
systematic uncertainty. 
Comparable uncertainties were  obtained by a dedicated
study~\cite{Bruneliere}. Uncertainties from  $\mathcal{O}(\alpha)$ corrections on the W-boson  decay angles are found to be negligible~\cite{dierckxsens}.
 
Uncertainties in the background cross sections 
and differential distributions are
possible sources of systematic effects. The cross sections of the
$\EE \ra \QQ (\gamma)$ and $\EE \ra$~ZZ$(\gamma)$ processes are varied 
within the theoretical uncertainty~\cite{MC-ffff} of $\pm 2\%$.
To reproduce the measured four-jet event rate of the 
$\EE \ra \QQ (\gamma)$~\cite{wwxsecpaper}, the corresponding Monte Carlo
is scaled by 12.7\%. Half of the effect is assigned as an additional 
systematic uncertainty.
Moreover, the $\cos\Theta_{\mathrm{W}^-}$ distributions for these 
backgrounds are reweighted with a linear function of slope $\pm 5\%$, 
in order to account for possible 
inaccuracies of the Monte Carlo predictions, giving a small additional
contribution to this systematic uncertainty. 

The uncertainties on the lepton and jet charge assignment are derived
from the statistical accuracy of the two data sets used to check the charge
measurement~\cite{villa,dierckxsens}: 
lepton-pair events in Z-peak calibration data for 
the measurement of the lepton charge
and  semi-leptonic W-pair events with muons 
for the charge of W bosons decaying into hadrons.
Uncertainties around 0.2\% are found for single 
tracks used for electron and tau reconstruction in the barrel and
between 1\% and 12\% in the endcaps, uncertainties around 0.06\% for the
charge of muons and
around 1.3\% for the charge of W bosons decaying into hadrons.

The agreement of data and Monte Carlo in the reconstruction of
angles and energies of jets and leptons is tested with di-jet and di-lepton
events collected during Z-peak calibration runs.
The uncertainties on scales and resolutions of
energy and angle measurements are propagated in the Monte Carlo and their
effect on the TGC results is assigned as a systematic uncertainty. 

The uncertainty caused by limited Monte Carlo statistics is evaluated 
by repeating the TGC fit with subsets of the total reference sample,
analysing the fit results as a function of the sample size and
extrapolating this shift to the full sample.

The modelling of initial-state radiation in KandY is included up
to $\mathcal{O}(\alpha^3)$ in the leading-logarithm approximation.  
The systematic uncertainty is estimated by comparing the fit results when
only ISR up to $\mathcal{O}(\alpha^2)$ is considered. 
A good description of final-state radiation (FSR) is important to 
properly reconstruct the phase space variables used in the TGC fit.
This effect is studied by repeating the TGC fit with Monte Carlo samples 
from which the events with FSR photons of energy above a 
cut-off, varied between $100~\MeV$ and $1~\GeV$, are removed. 

Systematic effects due to the uncertainty on the measurement
of the W mass and width are evaluated by varying these
parameters within the uncertainties of the world averages~\cite{PDG02}.

High statistics Monte Carlo samples generated with different hadronisation
schemes, PYTH\-IA~\cite{JETSET73}, HERWIG~\cite{HERWIG} and 
ARIADNE~\cite{ARIADNE}, are used to evaluate the effect of 
hadronisation modelling uncertainties. 
The average of the absolute value of the TGC shifts observed between
different models is assigned as systematic uncertainty.

Other final state phenomena which can influence the TGC fit are 
colour reconnection~\cite{FSI-CR} and 
Bose-Einstein~\cite{FSI-BE} effects.
Monte Carlo samples with implementation of 
different models of colour reconnection and Bose-Einstein
correlations are used to fit TGC's and
evaluate the associated systematic uncertainties by comparison with
the reference sample.
For colour reconnection the following models are tested: model
II~\cite{lonnblad} in ARIADNE, the scheme implemented
in HERWIG  and the SK~I~\cite{sjost} model with full reconnection
probability in PYTHIA.     
Based on a study of compatibility of SK~I with colour flow
between jets~\cite{l3-272}, only half the effect is considered.
The averages of the absolute values
of the shifts obtained using different models are 
quoted as systematic uncertainties. 
For Bose-Einstein correlation, the
LUBOEI~\cite{LUBOEI} BE$_{32}$ model as implemented in
PYTHIA with and without correlation between jets coming from
different W bosons is studied. The difference is taken as systematic
uncertainty. 

Systematic uncertainties for the single-W results are dominated by 
uncertainties on selection efficiencies and  
signal cross section~\cite{l3-252} and amount to 0.068 for $\kg$ and 0.08 for
$\Lg$.  

%%%%%%%%%%%%%%%%%%%%%%%%%%%%%%%%%%%%%%%%%%%%%%%%%%%%%%%%%%%%%%%%%%%%%%%%%%%%%%%%%%
%Results and Discussion
%%%%%%%%%%%%%%%%%%%%%%%%%%%%%%%%%%%%%%%%%%%%%%%%%%%%%%%%%%%%%%%%%%%%%%%%%%%%%%%%%%
\section{Results and Discussion}
\label{sec:tgcs}

The results obtained from the study of W-pair events collected at 
$\sqrt{s}= 189-209~\GeV$ are
combined taking into account correlations
of systematic errors between decay channels and between
data sets collected at different centre-of-mass energies. 

Further, they are combined with W-pair results obtained at lower
$\sqrt{s}$~\cite{l3-189}, with the single-W
results~\cite{l3-252} recalculated after removing the overlap with the W-pair
selection and with the results from single-photon events~\cite{singlePhoton}. 

\begin{figure}[p]
\begin{center}
  \epsfig{file=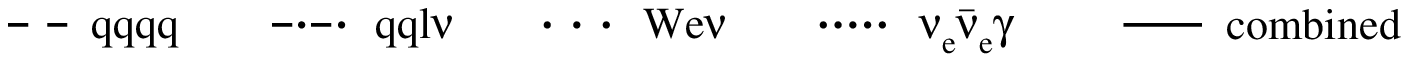,width=0.92\linewidth} \\ 
  \epsfig{file=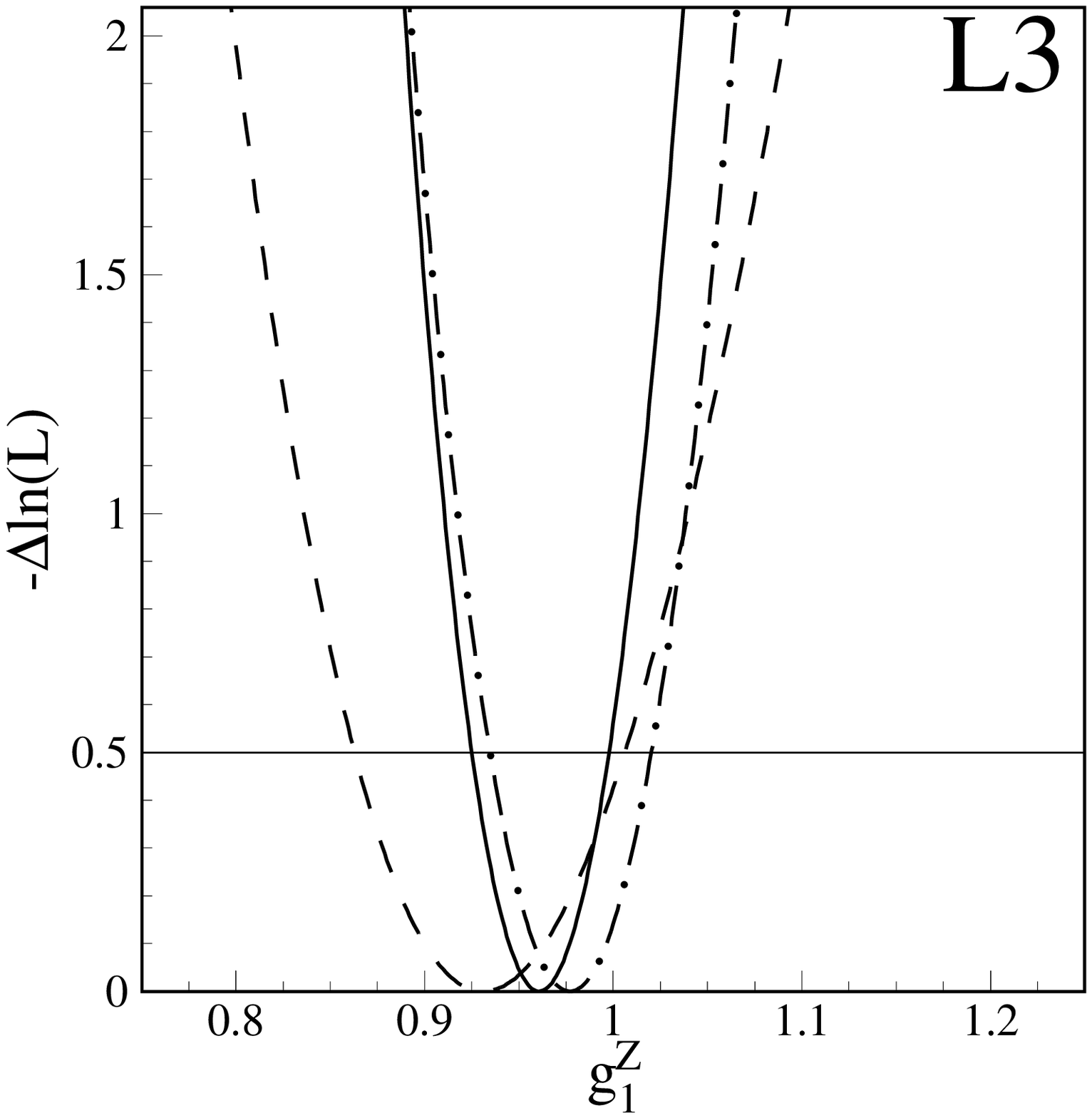,width=0.405\linewidth} \hspace*{1cm}
  \epsfig{file=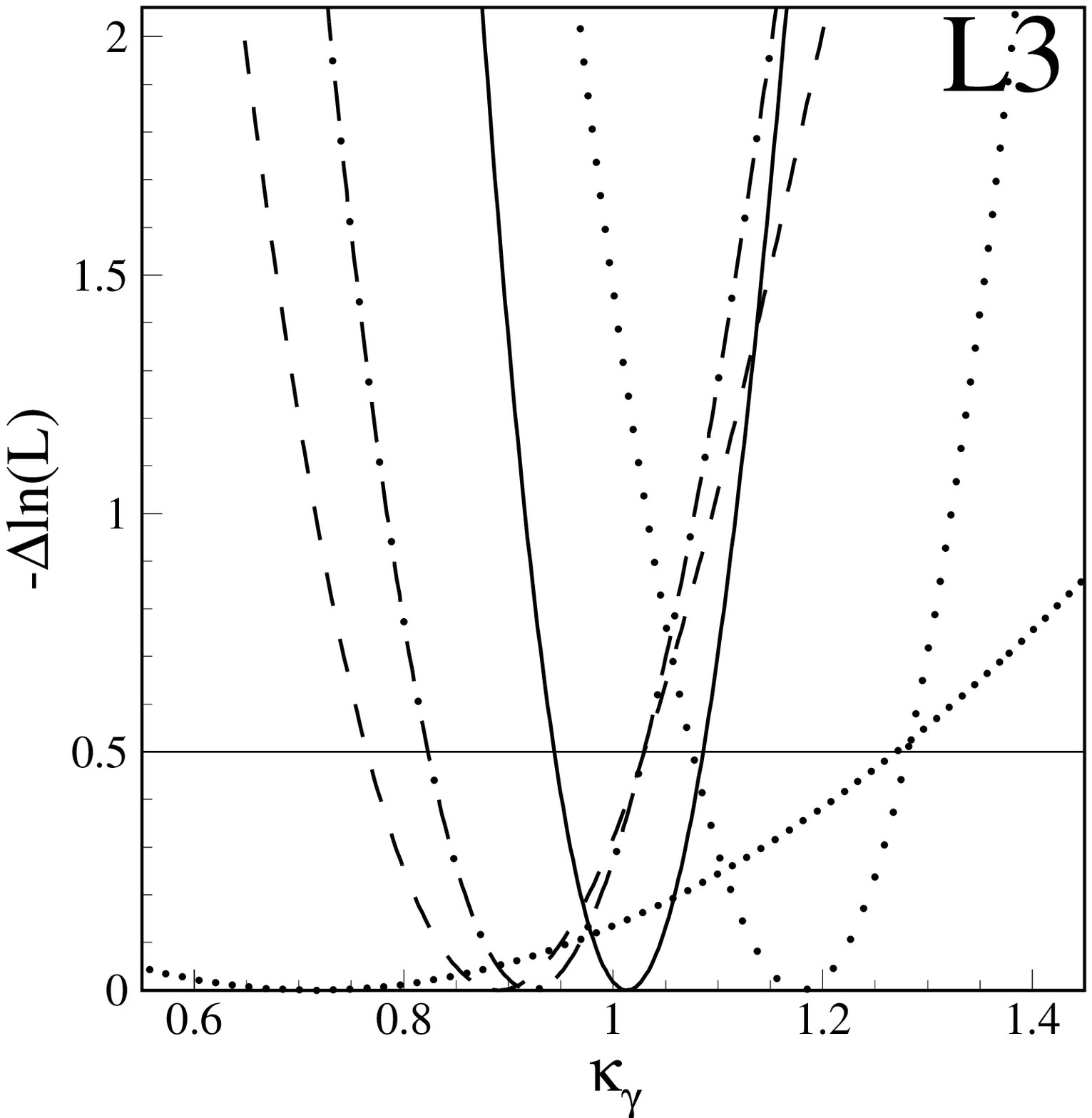,width=0.405\linewidth} \\ 
  \epsfig{file=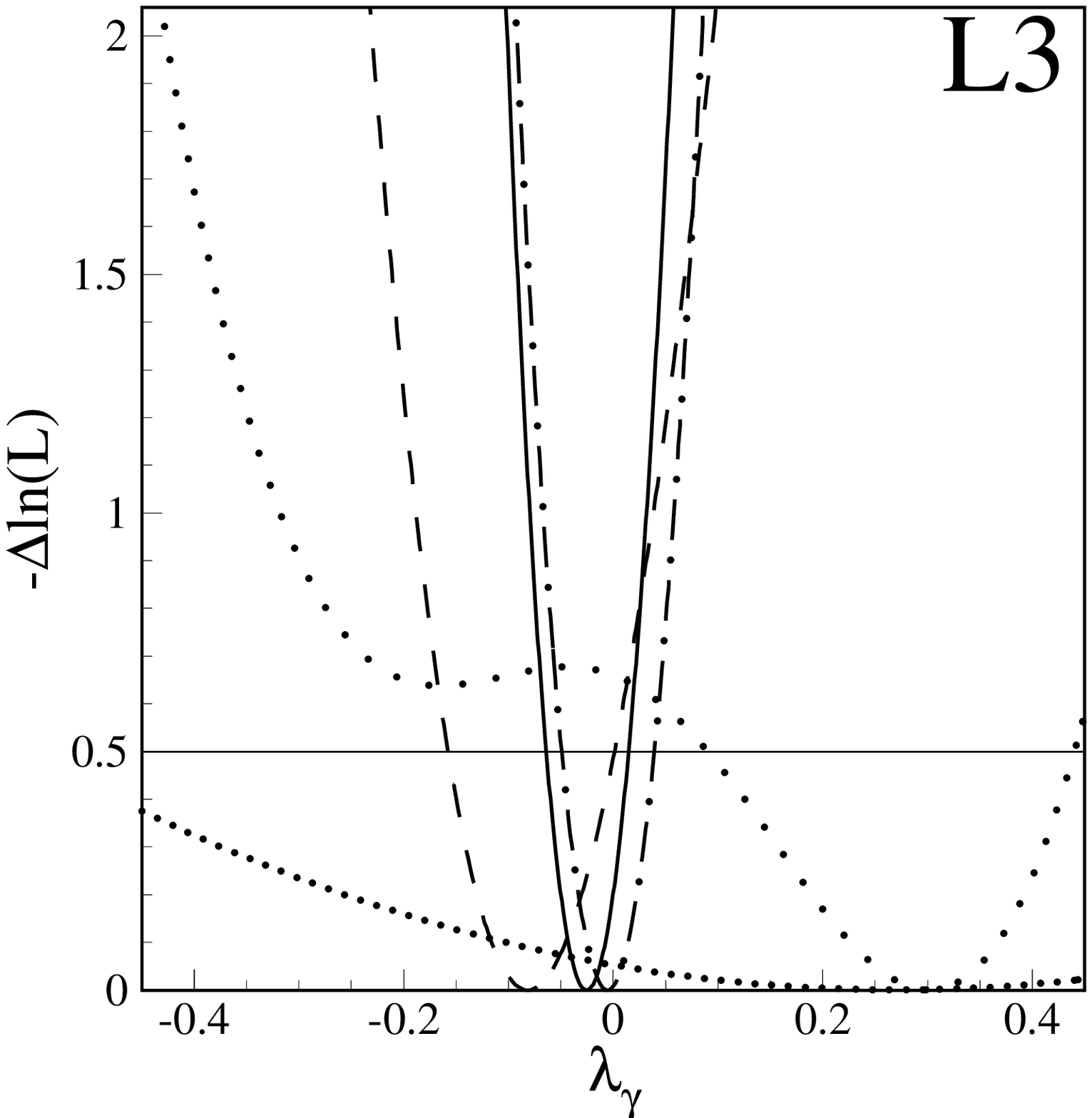,width=0.405\linewidth} \hspace*{1cm}
  \epsfig{file=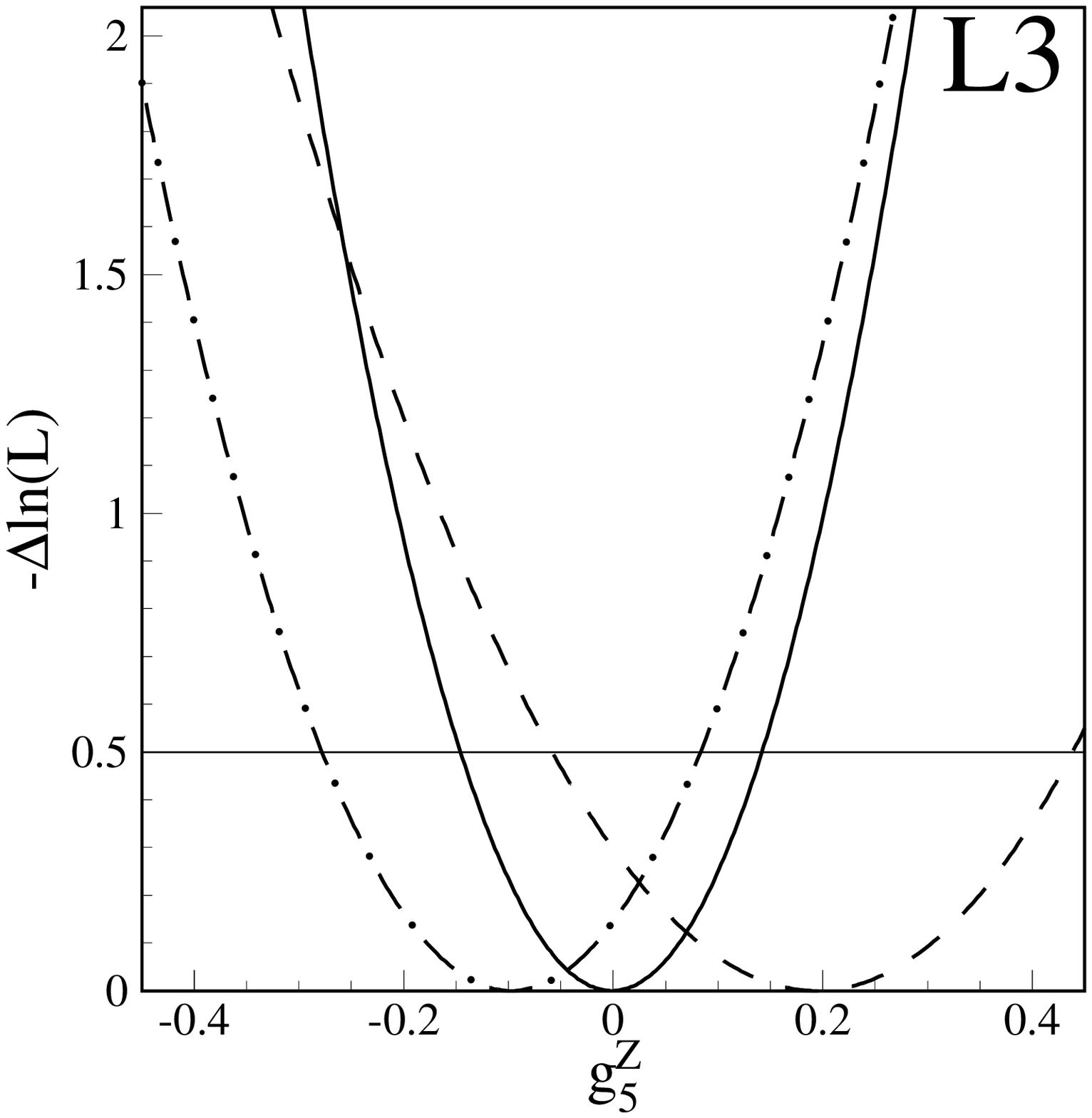,width=0.405\linewidth}  \\ 
  \epsfig{file=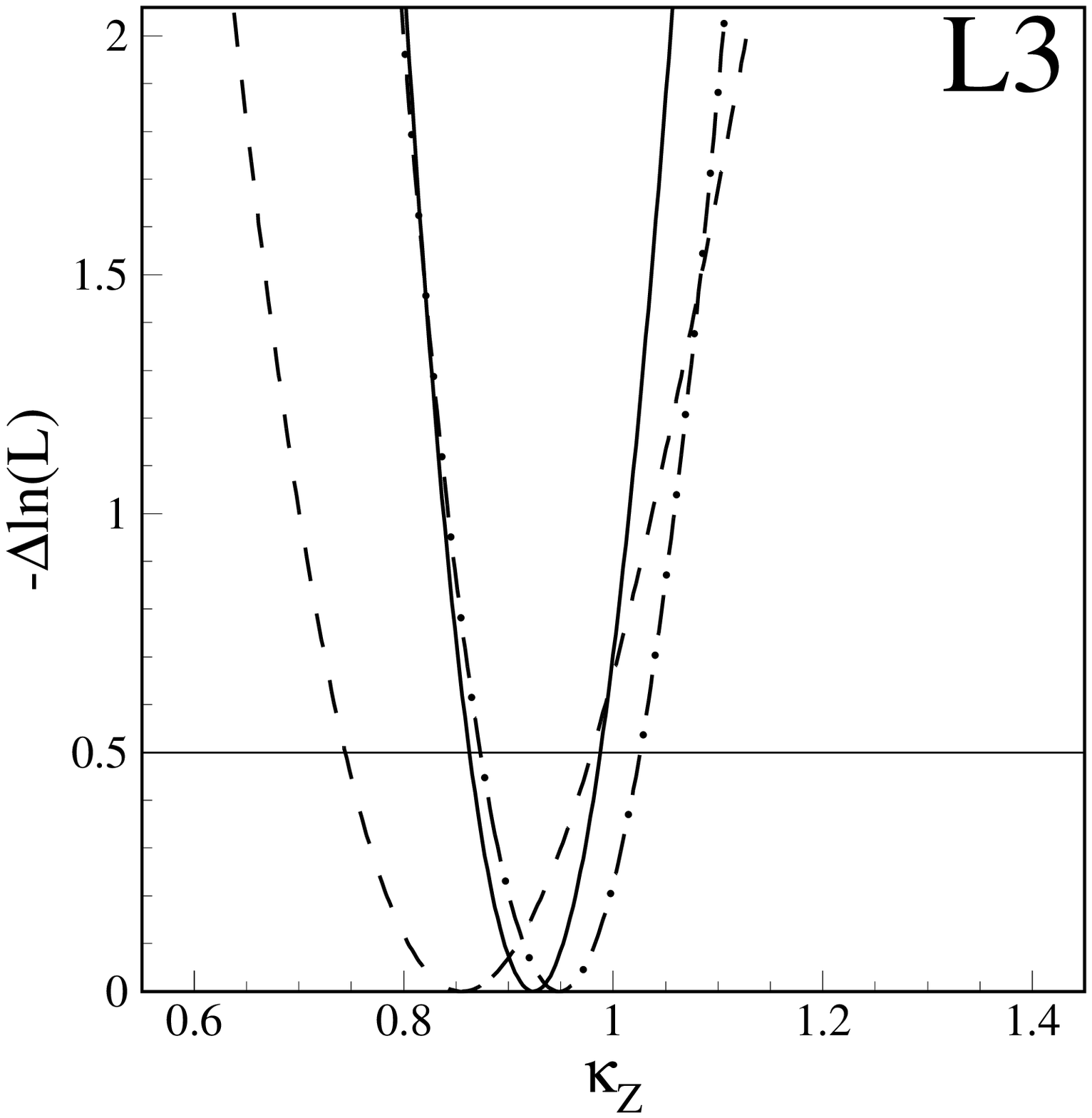,width=0.405\linewidth} \hspace*{1cm}
  \epsfig{file=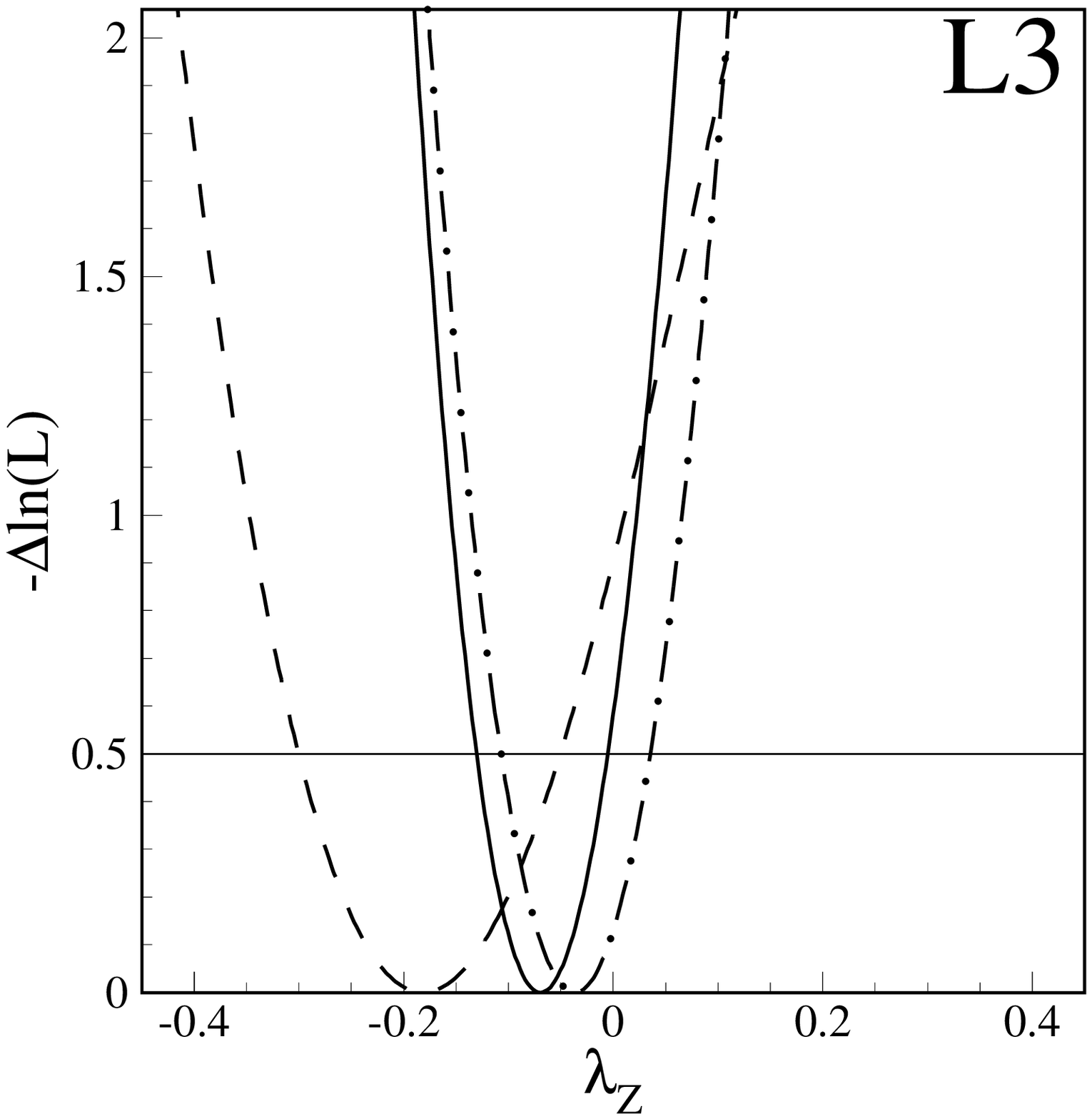,width=0.405\linewidth} 
  \caption[]{
    Change in negative log-likelihoods with respect to their minimum
    for one-parameter TGC fits. Systematic uncertainties are included.
    Contributions from different channels are indicated.
  }
\label{fig:logl1d}
\end{center}
\end{figure}

The results of one-parameter fits, in which only one coupling is
allowed to vary while the others are set to their Standard Model
values, are given in Table~\ref{tab:res1d1}.  
Negative log-likelihood curves are shown in Figure~\ref{fig:logl1d}.

\begin{figure}[p]
\begin{center}
\epsfig{file=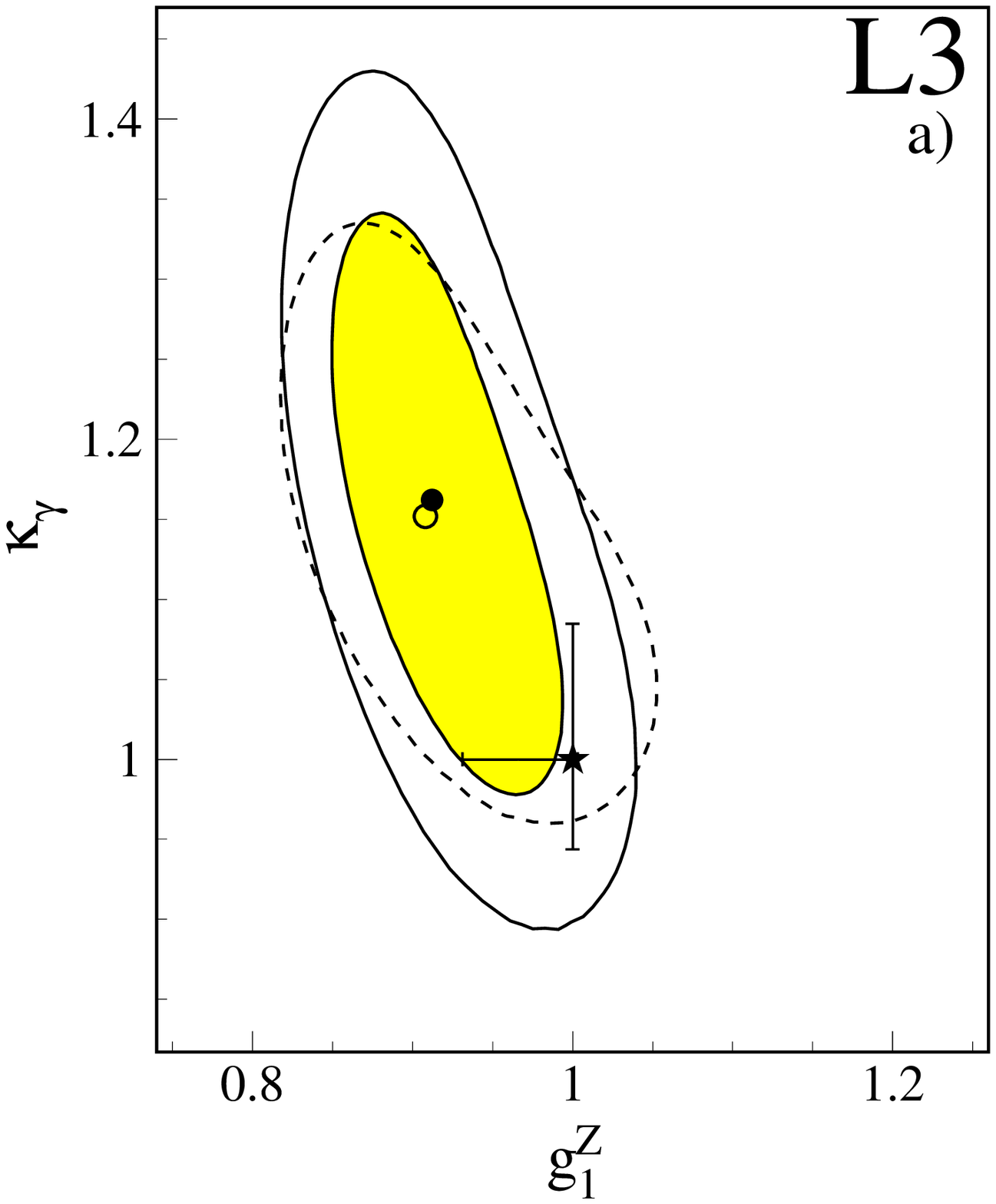,width=0.45\linewidth} \hspace*{1cm}
\epsfig{file=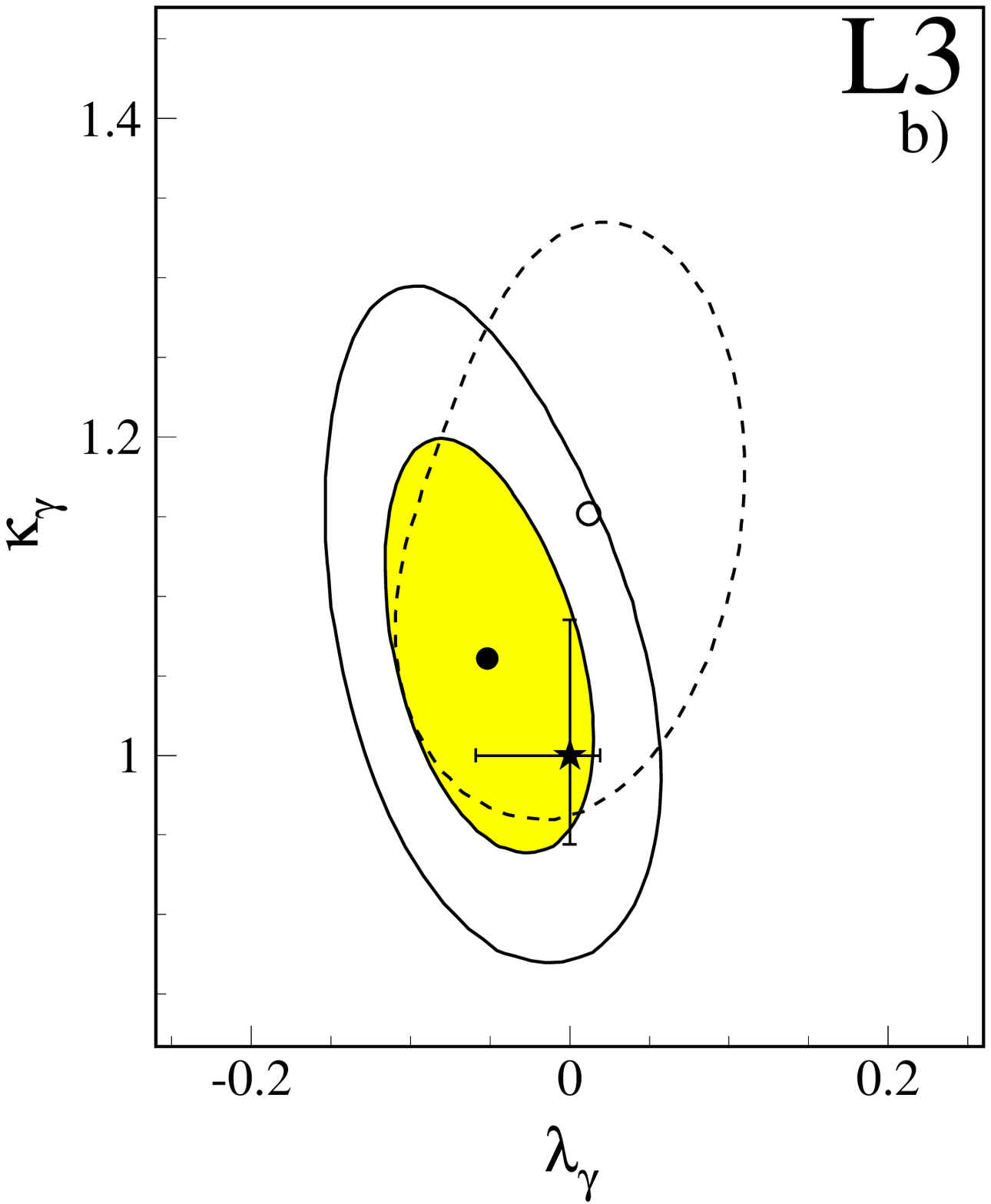,width=0.45\linewidth}\\
\epsfig{file=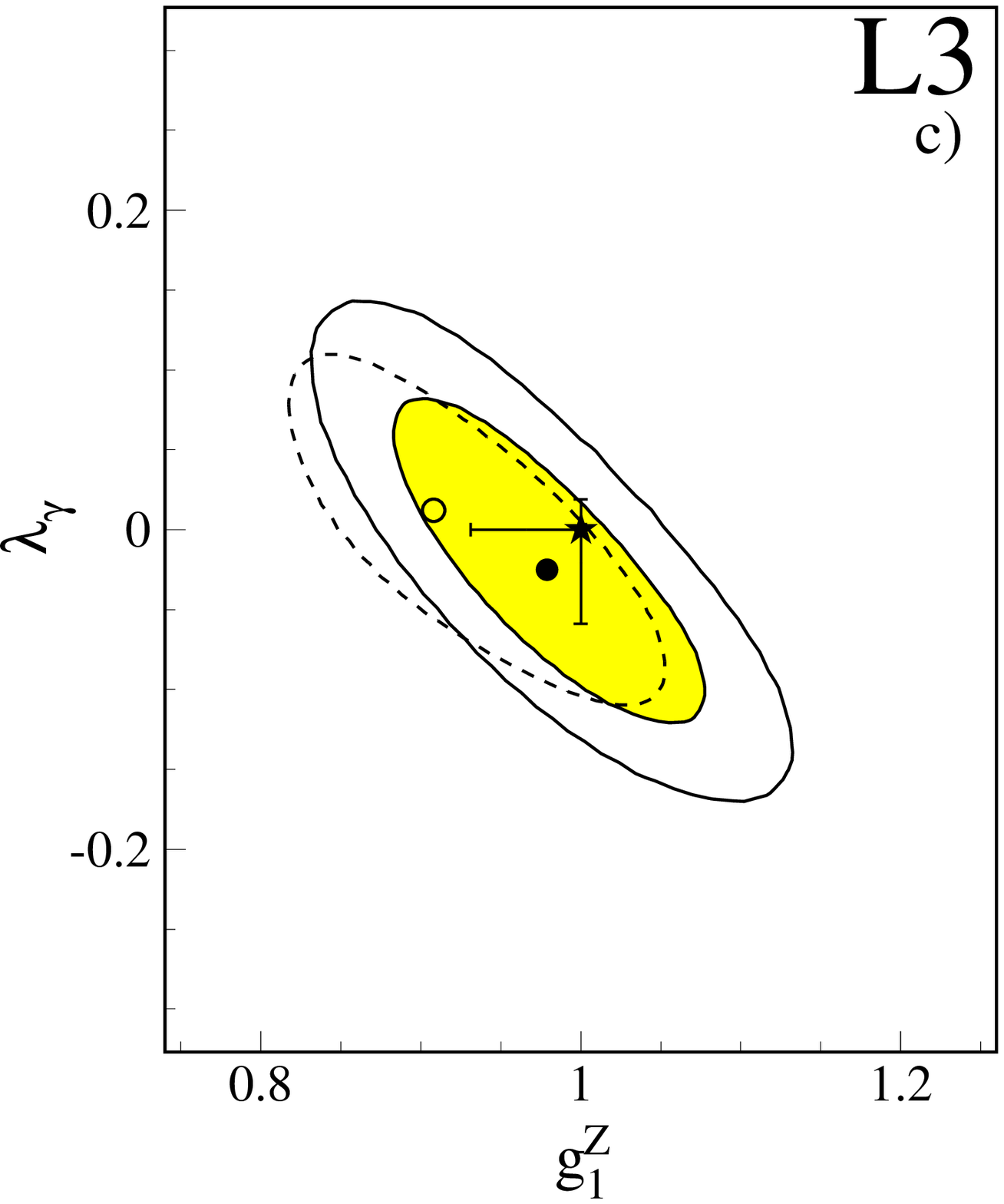,width=0.45\linewidth} \hspace*{1cm}
\epsfig{file=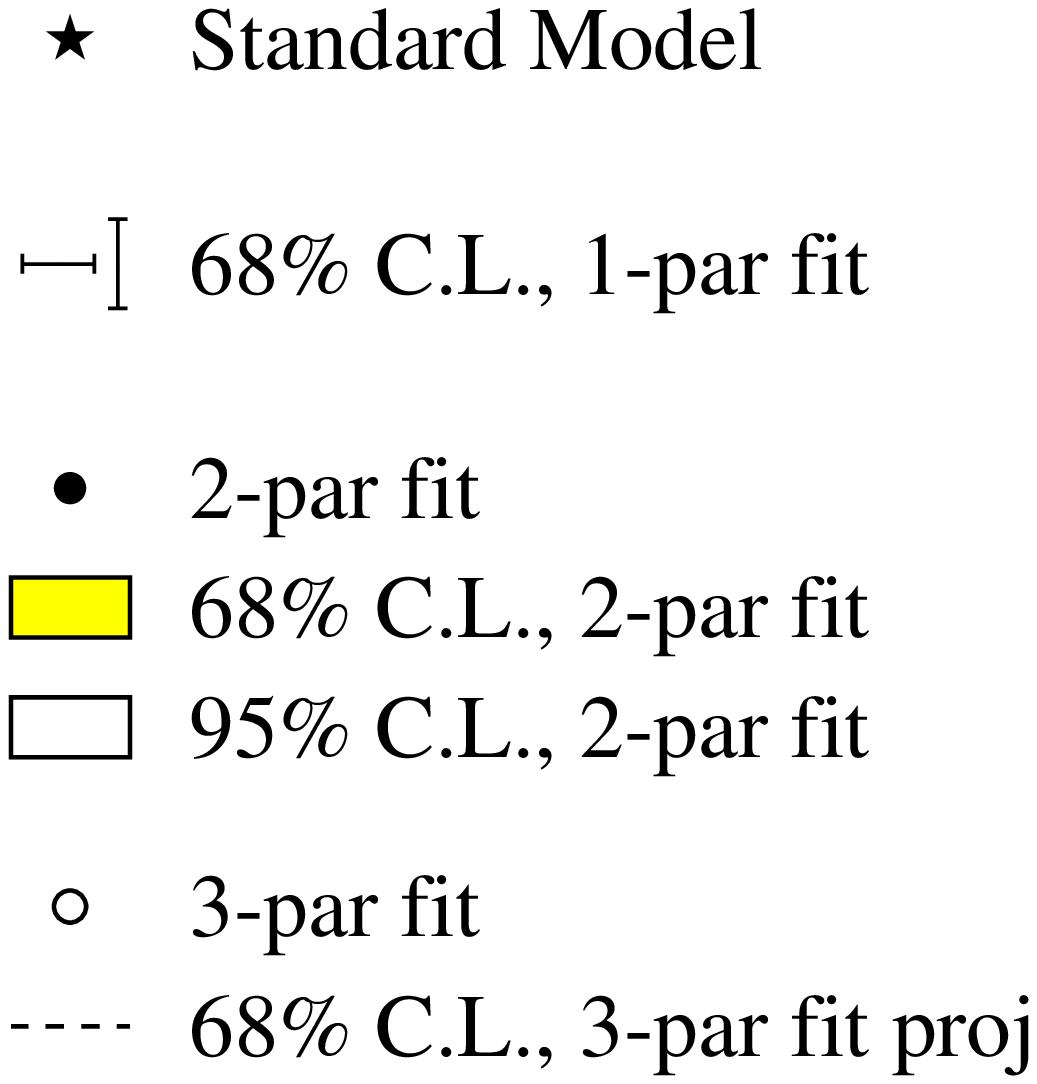,width=0.45\linewidth} 
\caption[]{Comparison of single- and multi-parameter TGC fits.
  The vertical and horizontal lines are the 68\% confidence level 
  intervals when all couplings but one are fixed to their 
  Standard Model values, indicated by a star.
  The shaded areas represent the 68\% confidence level regions
  for the two-parameter fits to the TGC's: a) $\giZ$ and $\kg$ with $\Lg=0$,
  b) $\Lg$ and $\kg$ with $\giZ=1$ and c) $\giZ$ and $\Lg$ with $\kg=1$.  
  The 95\% confidence level contours  are also given as solid lines.
  The dashed lines represent two-dimensional projections 
  of the three-parameter log-likelihoods.
  The constraints $\kZ = \giZ - \tan^2 \theta_{W} (\kg-1)$ and $\LZ =
  \Lg$ are imposed and all other couplings are set to their
  Standard Model values. Systematic uncertainties are included.
}
\label{fig:ac-ndlnl}
\end{center}
\end{figure}

Multi-parameter fits of TGC's allow a
model-independent  interpretation of the data.  
Fits to two of the couplings $\kg$, $\Lg$ and $\giZ$, 
keeping the third coupling fixed at its Standard Model value, are performed, 
as well as a simultaneous fit to all these couplings.
In each case the constraints 
$\kZ = \giZ - \tan^2 \theta_{W} (\kg-1)$ and $\LZ = 
\Lg$ are imposed. 
The results of these multi-parameter fits 
are reported in Table~\ref{tab:res2-3d}.
The contour curves of 68\% and 95\% confidence level for the two-parameter fits
are shown in Figure~\ref{fig:ac-ndlnl}.  
They correspond to a change
in the negative log-likelihood with respect to its minimum of $1.15$ and $3.00$,
respectively. 
Contours derived from three-parameter fits are also shown.
They are obtained requiring a log-likelihood 
change of $1.15$, but leaving the third coupling free to vary in the fit.
The comparison of the results derived from fits of different dimensionality
shows good agreement.

If the W boson were an extended object, \eg \ an ellipsoid of rotation with
longitudinal radius $a$ and transverse radius $b$, its size
and shape would be related to the TGCs by $R_{\PW} \equiv (a+b)/2 =
(\kg+\Lg-1)/m_{\mathrm W}$~\cite{Brodsky:1980} and $\Delta_{\PW} \equiv
(a^2-b^2)/2 = (5/4)(\kg-\Lg-1)/m_{\mathrm W}^2$~\cite{quad1}, 
where $m_{\mathrm W}$ is the mass of the W boson.   
The measurements show no evidence for the W boson to be an extended
object:
\begin{eqnarray}
     R_{\PW} & = & (0.3\pm1.9)\times 10^{-19}~\mathrm{m}   \\
\Delta_{\PW} & = & (0.89\pm0.83)\times 10^{-36}~\mathrm{m}^2 \,,
\end{eqnarray}
with a correlation coefficient of $-0.63$.

In conclusion, TGC's are measured with an accuracy of a few percent.
All single- and multi-parameter TGC results show good agreement with
the Standard Model expectation and confirm the existence of self-couplings among
the electroweak gauge bosons.  

\clearpage

%
%%%%%%%%%%%%%%%%%%%%%%%%%%%%%%%%%%%%%%%%%%%%%%%%%%%%%%%%%%%%%%%%%%%%%%%%%%%%%%%
% Bibliography
%%%%%%%%%%%%%%%%%%%%%%%%%%%%%%%%%%%%%%%%%%%%%%%%%%%%%%%%%%%%%%%%%%%%%%%%%%%%%%
\bibliographystyle{/afs/cern.ch/l3/paper/biblio/l3stylem}
\bibliography{tgc}

%
%%%%%%%%%%%%%%%%%%%%%%%%%%%%%%%%%%%%%%%%%%%%%%%%%%%%%%%%%%%%%%%%%%%%%%%%%%%%%%
% Author List
%%%%%%%%%%%%%%%%%%%%%%%%%%%%%%%%%%%%%%%%%%%%%%%%%%%%%%%%%%%%%%%%%%%%%%%%%%%%%%
%
\newpage
\typeout{   }     
\typeout{Using author list for paper 281 -  }
\typeout{$Modified: Jul 15 2001 by smele $}
\typeout{!!!!  This should only be used with document option a4p!!!!}
\typeout{   }
%
%
%
%  L A T E X  version!!
%
%
% Make sure that the Lep package has been used!
%\input{Lep.sty}%
%
%\ifx\LepCalled\undefined%
%\typeout{     }%
%\typeout{!!!!!!!!!!!!!!!!!!!!!!!!!!!!!!!!!!!!!!!!!!!!!!!!!!!!!!!!!!!}%
%\typeout{Yikes.  You haven't used the Lep package!}%
%\typeout{Please put \protect\usepackage\protect{Lep\protect} in your preamble,
%         followed by}%
%\typeout{\protect\Lep\protect{1\protect} or \protect\Lep\protect{2\protect}}%
%\typeout{     }%
%\typeout{For now you will get a Lep phase 2 authorlist (may not be right!).}%
%\typeout{!!!!!!!!!!!!!!!!!!!!!!!!!!!!!!!!!!!!!!!!!!!!!!!!!!!!!!!!!!!}%
%\typeout{     }%
%\Lep{2}\fi%

\newcount\tutecount  \tutecount=0
\def\tutenum#1{\global\advance\tutecount by 1 \xdef#1{\the\tutecount}}
\def\tute#1{$^{#1}$}
\tutenum\aachen            % 1
\tutenum\nikhef            % 2
\tutenum\mich              % 3
\tutenum\lapp              % 4
\tutenum\basel             % 5
\tutenum\lsu               % 6
\tutenum\beijing           % 7
\tutenum\bologna           % 8
\tutenum\tata              % 9 
\tutenum\ne                % 10
\tutenum\bucharest         % 11
\tutenum\budapest          % 12
\tutenum\mit               % 13
\tutenum\panjab            % 14 
\tutenum\debrecen          % 15
\tutenum\dublin            % 16
\tutenum\florence          % 17
\tutenum\cern              % 18
\tutenum\wl                % 19
\tutenum\geneva            % 20
\tutenum\hefei             % 21
\tutenum\lausanne          % 22
\tutenum\lyon              % 23
\tutenum\madrid            % 24
\tutenum\florida           % 25
\tutenum\milan             % 26
\tutenum\moscow            % 27
\tutenum\naples            % 29
\tutenum\cyprus            % 30
\tutenum\nymegen           % 31
\tutenum\caltech           % 32
\tutenum\perugia           % 33
\tutenum\peters            % 34
\tutenum\cmu               % 35
\tutenum\potenza           % 36
\tutenum\prince            % 37
\tutenum\riverside         % 38
\tutenum\rome              % 39
\tutenum\salerno           % 40
\tutenum\ucsd              % 41
\tutenum\sofia             % 42
\tutenum\korea             % 43
\tutenum\purdue            % 44
\tutenum\psinst            % 45
\tutenum\zeuthen           % 46
\tutenum\eth               % 47
\tutenum\hamburg           % 48
\tutenum\taiwan            % 49
\tutenum\tsinghua          % 50

{
\parskip=0pt
\noindent
{\bf The L3 Collaboration:}
\ifx\selectfont\undefined%  old style font selection
 \baselineskip=10.8pt
 \baselineskip\baselinestretch\baselineskip
 \normalbaselineskip\baselineskip
 \ixpt
\else%                      new style font selection
 \fontsize{9}{10.8pt}\selectfont
\fi
\medskip
\tolerance=10000
\hbadness=5000
\raggedright
\hsize=162truemm\hoffset=0mm
\def\r{\rlap,}
\noindent

P.Achard\r\tute\geneva\ 
O.Adriani\r\tute{\florence}\ 
M.Aguilar-Benitez\r\tute\madrid\ 
J.Alcaraz\r\tute{\madrid}\ 
G.Alemanni\r\tute\lausanne\
J.Allaby\r\tute\cern\
A.Aloisio\r\tute\naples\ 
M.G.Alviggi\r\tute\naples\
H.Anderhub\r\tute\eth\ 
V.P.Andreev\r\tute{\lsu,\peters}\
% ...
\newpage
%\rule{\textwidth}{0.4pt}
\begin{list}{A}{\itemsep=0pt plus 0pt minus 0pt\parsep=0pt plus 0pt minus 0pt
                \topsep=0pt plus 0pt minus 0pt}
\item[\aachen]
 III. Physikalisches Institut, RWTH, D-52056 Aachen, Germany$^{\S}$
\item[\nikhef] National Institute for High Energy Physics, NIKHEF, 
     and University of Amsterdam, NL-1009 DB Amsterdam, The Netherlands
\item[\mich] University of Michigan, Ann Arbor, MI 48109, USA
\item[\lapp] Laboratoire d'Annecy-le-Vieux de Physique des Particules, 
     LAPP,IN2P3-CNRS, BP 110, F-74941 Annecy-le-Vieux CEDEX, France
\item[\basel] Institute of Physics, University of Basel, CH-4056 Basel,
     Switzerland
\item[\lsu] Louisiana State University, Baton Rouge, LA 70803, USA
\item[\beijing] Institute of High Energy Physics, IHEP, 
  100039 Beijing, China$^{\triangle}$ 
\item[\bologna] University of Bologna and INFN-Sezione di Bologna, 
     I-40126 Bologna, Italy
\item[\tata] Tata Institute of Fundamental Research, Mumbai (Bombay) 400 005, India
\item[\ne] Northeastern University, Boston, MA 02115, USA
\item[\bucharest] Institute of Atomic Physics and University of Bucharest,
     R-76900 Bucharest, Romania
\item[\budapest] Central Research Institute for Physics of the 
     Hungarian Academy of Sciences, H-1525 Budapest 114, Hungary$^{\ddag}$
\item[\mit] Massachusetts Institute of Technology, Cambridge, MA 02139, USA
\item[\panjab] Panjab University, Chandigarh 160 014, India.
\item[\debrecen] KLTE-ATOMKI, H-4010 Debrecen, Hungary$^\P$
\item[\dublin] Department of Experimental Physics,
  University College Dublin, Belfield, Dublin 4, Ireland
\item[\florence] INFN Sezione di Firenze and University of Florence, 
     I-50125 Florence, Italy
\item[\cern] European Laboratory for Particle Physics, CERN, 
     CH-1211 Geneva 23, Switzerland
\item[\wl] World Laboratory, FBLJA  Project, CH-1211 Geneva 23, Switzerland
\item[\geneva] University of Geneva, CH-1211 Geneva 4, Switzerland
\item[\hefei] Chinese University of Science and Technology, USTC,
      Hefei, Anhui 230 029, China$^{\triangle}$
\item[\lausanne] University of Lausanne, CH-1015 Lausanne, Switzerland
\item[\lyon] Institut de Physique Nucl\'eaire de Lyon, 
     IN2P3-CNRS,Universit\'e Claude Bernard, 
     F-69622 Villeurbanne, France
\item[\madrid] Centro de Investigaciones Energ{\'e}ticas, 
     Medioambientales y Tecnol\'ogicas, CIEMAT, E-28040 Madrid,
     Spain${\flat}$ 
\item[\florida] Florida Institute of Technology, Melbourne, FL 32901, USA
\item[\milan] INFN-Sezione di Milano, I-20133 Milan, Italy
\item[\moscow] Institute of Theoretical and Experimental Physics, ITEP, 
     Moscow, Russia
\item[\naples] INFN-Sezione di Napoli and University of Naples, 
     I-80125 Naples, Italy
\item[\cyprus] Department of Physics, University of Cyprus,
     Nicosia, Cyprus
\item[\nymegen] University of Nijmegen and NIKHEF, 
     NL-6525 ED Nijmegen, The Netherlands
\item[\caltech] California Institute of Technology, Pasadena, CA 91125, USA
\item[\perugia] INFN-Sezione di Perugia and Universit\`a Degli 
     Studi di Perugia, I-06100 Perugia, Italy   
\item[\peters] Nuclear Physics Institute, St. Petersburg, Russia
\item[\cmu] Carnegie Mellon University, Pittsburgh, PA 15213, USA
\item[\potenza] INFN-Sezione di Napoli and University of Potenza, 
     I-85100 Potenza, Italy
\item[\prince] Princeton University, Princeton, NJ 08544, USA
\item[\riverside] University of Californa, Riverside, CA 92521, USA
\item[\rome] INFN-Sezione di Roma and University of Rome, ``La Sapienza",
     I-00185 Rome, Italy
\item[\salerno] University and INFN, Salerno, I-84100 Salerno, Italy
\item[\ucsd] University of California, San Diego, CA 92093, USA
\item[\sofia] Bulgarian Academy of Sciences, Central Lab.~of 
     Mechatronics and Instrumentation, BU-1113 Sofia, Bulgaria
\item[\korea]  The Center for High Energy Physics, 
     Kyungpook National University, 702-701 Taegu, Republic of Korea
\item[\purdue] Purdue University, West Lafayette, IN 47907, USA
\item[\psinst] Paul Scherrer Institut, PSI, CH-5232 Villigen, Switzerland
\item[\zeuthen] DESY, D-15738 Zeuthen, Germany
\item[\eth] Eidgen\"ossische Technische Hochschule, ETH Z\"urich,
     CH-8093 Z\"urich, Switzerland
\item[\hamburg] University of Hamburg, D-22761 Hamburg, Germany
\item[\taiwan] National Central University, Chung-Li, Taiwan, China
\item[\tsinghua] Department of Physics, National Tsing Hua University,
      Taiwan, China
\item[\S]  Supported by the German Bundesministerium 
        f\"ur Bildung, Wissenschaft, Forschung und Technologie
\item[\ddag] Supported by the Hungarian OTKA fund under contract
numbers T019181, F023259 and T037350.
\item[\P] Also supported by the Hungarian OTKA fund under contract
  number T026178.
\item[$\flat$] Supported also by the Comisi\'on Interministerial de Ciencia y 
        Tecnolog{\'\i}a.
\item[$\sharp$] Also supported by CONICET and Universidad Nacional de La Plata,
        CC 67, 1900 La Plata, Argentina.
\item[$\triangle$] Supported by the National Natural Science
  Foundation of China.
\end{list}
}
\vfill

%%% Local Variables: 
%%% mode: latex
%%% TeX-master: t
%%% End:

\clearpage

%%%%%%%%%%%%%%%%%%%%%%%%%%%%%%%%%%%%%%%%%%%%%%%%%%%%%%%%%%%%%%%%%%%%%%%%%%%%%%
%                         TABLES
%%%%%%%%%%%%%%%%%%%%%%%%%%%%%%%%%%%%%%%%%%%%%%%%%%%%%%%%%%%%%%%%%%%%%%%%%%%%%%

\begin{table}[p]
  \begin{center}
    \renewcommand{\arraystretch}{1.2}
    \begin{tabular}{|c| r r r r r r r r|}
      \hline
      $\langle \RS \rangle \; [\GeV]$ & 
      188.6 & 191.6 & 195.5 & 199.6 & 201.8 & 204.8 & 206.5 & 208.0 \\  
      $\mathcal{L} \; [\mathrm{pb}^{-1}]$ &
      176.8 &  29.8 &  84.1 &  83.3 &  37.1 &  79.0 & 130.5 &   8.6 \\   
\hline
\end{tabular}
\caption[]{The average centre-of-mass energies, $\langle \RS \, \rangle$,
  and total integrated luminosities, $\mathcal{L}$, used for the W-pair
  analysis.}
\label{tab:luminosities}
\end{center}
\end{table}

\begin{table}[p]
\begin{center}
\renewcommand{\arraystretch}{1.2}
\begin{tabular}{| l | c | c |}
\hline
Process & $N_{\mathrm{data}}$ & $N_{\mathrm{bg}}$  \\
\hline                                                        
$\WW\rightarrow\LNLN$          & $ \pz 207$ & $\pz \pz 28.1$  \\
$\WW\rightarrow\QQEN$          & $1263$ & $\pz 118.1$  \\
$\WW\rightarrow\QQMN$          & $1187$ & $\pz 118.0$  \\
$\WW\rightarrow\QQTN$          & $1017$ & $\pz 348.4$  \\
$\WW\rightarrow\QQQQ$          & $5219$    & $1109.2$ \\
\hline
$\PW\EN,~\PW\rightarrow\LN$ & $ \pz 121$ & $\pz \pz 10.4$  \\
$\PW\EN,~\PW\rightarrow\QQ$ & $ \pz 584$ & $\pz 342.2$  \\
\hline
\end{tabular}
\caption[]{
  Numbers of selected data events, $N_{\mathrm{data}}$, and expected
  background events, $N_{\mathrm{bg}}$, for the W-pair analysis
  at $\RS= 189 - 209~\GeV$
  and for the single-W analysis at $\RS= 161 - 209~\GeV$. } 
\label{tab:events-all}
\end{center}
\end{table}

\begin{table}[p]
\begin{center}
\renewcommand{\arraystretch}{1.2}
\begin{tabular}{|l| r| r| r| r| r| r|}
\hline
\multicolumn{1}{|c|}{Source}& 
\multicolumn{6}{c|}{Systematic uncertainty} \\ \cline{2-7}  
\multicolumn{1}{|c|}{of uncertainty}&  
\multicolumn{1}{c|}{$\; \giZ$} & \multicolumn{1}{c|}{$\kg$}   &  
\multicolumn{1}{c|}{$\Lg$}     & \multicolumn{1}{c|}{$\; \gvZ$} &
\multicolumn{1}{c|}{$\kZ$}     & \multicolumn{1}{c|}{$\LZ$}     \\
\hline
Uncertainty on $\sigma_{\PW\PW}$   &  0.003  &  0.018  &  0.006
                                 &  0.03   &  0.009  &  0.014  \\
$\cal{O}(\alpha)$ corrections  on  $\cos\Theta_{\mathrm{W}^-}$  &  0.004  &  0.004  &  0.003 
                                 &  0.01   &  0.011  &  0.007   \\
Background modelling             &  0.005  &  0.019  &  0.006
                                 &  0.02   &  0.009  &  0.014   \\
Jet charge confusion             &  0.001   &  0.006  &  0.002 
                                 &  $<0.01$ &  0.002  &  0.005   \\

Lepton charge confusion          &  0.003  &  0.013  &  0.007 
                                 &  $\pz0.01$ &  0.005  &  0.009   \\
Jet and lepton measurement       &  0.001  &  0.003  &  0.002    
                                 &  0.01   &  0.002  &  0.004   \\
Monte Carlo statistics           &  0.012  &  0.010  &  0.014
                                 &  0.02   &  0.016  &  0.007    \\
ISR and FSR                      &  0.001  &  0.016  &  0.001
                                 &  0.01   &  0.002  &  0.002   \\
W mass and width                 &  0.001  &  0.005  &  0.002    
                                 &  0.01   &  0.002  &  0.004   \\
Fragmentation                    &  0.003  &  0.002  &  0.001
                                 &  0.02   &  0.004  &  0.001    \\
Bose Einstein correlations       &  0.001  &  0.001  &  0.001  
                                 &  $<0.01$  &  0.001 &  0.003  \\
% %
Colour reconnection              &  0.001 &  0.004   &  0.001 
                                 &  0.02  &  0.002  &   0.003   \\
% %
\hline
Total systematic uncertainty     &  0.015  &  0.039  & 0.017
                                 &  0.05   &  0.024  & 0.023    \\
\hline
\end{tabular}
\caption[]{Systematic uncertainties on TGC's determined from semi-leptonic
 and hadronic W-pairs.
 For each coupling the uncertainties are obtained in one-parameter fits, by setting all other
 couplings  to their Standard Model values. The constraints 
$\kZ = \giZ - \tan^2 \theta_{W} (\kg-1)$ 
and $\LZ = \Lg$  are imposed on the first three couplings.
}
\label{tab:sysww}
\end{center}
\end{table}

\begin{table}[p]
\begin{center}
\renewcommand{\arraystretch}{1.3}
\begin{tabular}{|c|c|c|c|}
\hline
Coupling   &    $\giZ$   &    $\kg$    &    $\Lg$      \\
\hline
% %
$\nu_e \bar{\nu}_e\gamma$ 189--209 $\GeV$ &   &
                     $0.7\pm 0.5 \pm 0.3$ &
                     $0.3\pm 0.7 \pm 0.4$ \\
We$\nu$ ~161--209 $\GeV$&   & $1.179\amp{0.080}{0.076}\pm 0.068$ &          
                        $0.30\amp{0.19}{0.11}\pm 0.08$   \\
WW ~161--209 $\GeV$  & $0.966\amp{0.032}{0.034}\pm 0.015$ & 
                     $0.910\amp{0.066}{0.074}\pm 0.039$ &
                     $-0.024\amp{0.033}{0.035}\pm 0.017$ \\
\hline
All channels combined  & $0.966\amp{0.032}{0.034} \pm 0.015$ & 
                         $1.013\amp{0.064}{0.067} \pm 0.026$  &
                         $-0.021\amp{0.034}{0.035} \pm 0.017$ \\
Standard Model value            &  1.0    &   1.0    & 0.0     \\
\hline
\multicolumn{4}{}{} \\
\hline
Coupling &        $\gvZ$        &     $\kZ$       &         $\LZ$         \\
\hline
WW 189-209 $\GeV$    & $0.00\pm0.13 \pm 0.05$ &
                       $0.924\amp{0.056}{0.059}  \pm 0.024$ &
                       $-0.088\amp{0.057}{0.060} \pm 0.023$ \\
Standard Model value          & 0.0     &  1.0    &   0.0    \\
\hline
\end{tabular}
\caption{Results of one-parameter fits to the TGC's $\giZ$, $\kg$,
  $\Lg$, $\gvZ$, $\kZ$ and $\LZ$ based on single-photon events,
  single-W  events and hadronic and semi-leptonic
  W-pairs, and  their combination. The single-W
  results are obtained after removing events 
  selected as W-pair. 
  All results are at 68\% confidence level.
  For each TGC fit, all other parameters are set to their Standard Model values;
  for the set  $\giZ$, $\kg$ and $\Lg$ the constraints $\kZ = \giZ -
  \tan^2 \theta_{W} (\kg-1)$ and $\LZ = \Lg$ are imposed.  The first
  uncertainty is statistical, the second systematic.}
\label{tab:res1d1}
\end{center}
\end{table}

\begin{table}[p]
\begin{center}

\renewcommand{\arraystretch}{1.35}
\begin{tabular}{|c|c|c|c|ccc|}
\hline
Fit       & Standard     & \multicolumn{2}{|c|}{Results}   &
\multicolumn{3}{|c|}{correlation coeffs.}\\  \cline{3-7}
parameter & Model & \hspace{4mm}(68\% CL)\hspace{4mm} 
&  (95\% CL)  & $\giZ$  & $\kg$   & $\Lg$   \\ 
\hline
 \multicolumn{7}{|c|}{two-parameter fits} \\
 \hline
 $\giZ$ & 1.0 & $\phantom{+}0.912^{+0.054}_{-0.044}$ & 
$ [\phantom{+}0.83, 1.02]$   & $\phantom{+}1.00$ & $-0.71$ &  \\
 $\kg$  & 1.0 & $\phantom{+}1.162^{+0.124}_{-0.129}$ & 
$ [\phantom{+}0.94, 1.38]$   &        & $\phantom{+}1.00$  &  \\
 \hline
 $\kg$ & 1.0 &  $\phantom{+}1.061^{+0.089}_{-0.082}$ & 
 $ [\phantom{+}0.91, 1.24]$   &     & $\phantom{+}1.00$  & $-0.42$  \\
 $\Lg$ & 0.0 & $-0.052^{+0.044}_{-0.042}$ &  
 $ [-0.13, 0.03]$     &     &    &   $ \phantom{+}1.00$     \\
 \hline
 $\giZ$ & 1.0 & $\phantom{+}0.979^{+0.066}_{-0.065}$ &  
 $ [\phantom{+}0.86, 1.10]$  & $\phantom{+}1.00$ &     & $-0.82$ \\
 $\Lg$  & 0.0 &  $-0.025^{+0.071}_{-0.065}$ & 
 $ [-0.14, 0.11]$   &        &     & $\phantom{+}1.00$    \\
 \hline
 \multicolumn{7}{|c|}{three-parameter fit} \\
 \hline
 $\giZ$ & 1.0 &  $\phantom{+}0.91^{+0.10}_{-0.07}$ & 
 $ [\phantom{+}0.80, 1.08]$ & $\phantom{+}1.00$ & $-0.74$ & $-0.80$ \\
 $\kg$  & 1.0 &  $\phantom{+}1.15^{+0.13}_{-0.14}$ & 
 $ [\phantom{+}0.92, 1.38]$ &        & $\phantom{+} 1.00$ & $ \phantom{+}0.44$ \\
 $\Lg$  & 0.0 &  $\phantom{+}0.01^{+0.07}_{-0.08}$ &  
 $ [-0.14, 0.14]$   &        &         & $\phantom{+}1.00$ \\
\hline 
\end{tabular}
\caption[]{
  Results of  two- and three-parameter fits of the couplings
  $\kg$, $\Lg$ and $\giZ$ with the constraints
  $\kZ = \giZ - \tan^2 \theta_{W} (\kg-1)$ and $\LZ = \Lg$; 
  all other couplings are set to their Standard Model values. 
  Correlation coefficients are also shown.  
  Systematic uncertainties are included.}
\label{tab:res2-3d}
\end{center}
\end{table}

\clearpage

%%%%%%%%%%%%%%%%%%%%%%%%%%%%%%%%%%%%%%%%%%%%%%%%%%%%%%%%%%%%%%%%%%%%%%%%%%%%%%
%                       FIGURES
%%%%%%%%%%%%%%%%%%%%%%%%%%%%%%%%%%%%%%%%%%%%%%%%%%%%%%%%%%%%%%%%%%%%%%%%%%%%%%

\end{document}